%% file: main.tex
\documentclass[1p]{elsarticle}

\pdfoutput=1

\usepackage[english]{babel}	
\usepackage{amsmath} 
\usepackage{amsfonts} 
\usepackage{csquotes}
\usepackage{multicol}
\usepackage{multirow}
\usepackage{float}
\usepackage{booktabs,array,dcolumn}
\usepackage[colorlinks=true,linkcolor=blue]{hyperref}%

\newcommand{\VMS}{\textrm{VMS}}

\newcommand{\EVM}{\textrm{EVM}}

\newcommand{\lr}[1]{\left(#1\right)} 
\newcommand{\pderiv}[2]{\frac{\partial #1}{\partial #2}}

\newcommand{\bld}[1]{\mathbf{#1}}
\newcommand{\bldsy}[1]{\boldsymbol{#1}}

\newcommand{\ub}{\bld{u}}
\newcommand{\Bb}{\bld{B}}

\newcommand{\mom}{\textrm{V}}
\newcommand{\ind}{\textrm{I}}
\newcommand{\tot}{\textrm{T}}

\newcommand{\nlv}{\bldsy{\mathcal{N}}^{\mom}}
\newcommand{\nli}{\bldsy{\mathcal{N}}^{\ind}}

\newcommand{\V}{\bldsy{\mathcal{V}}}
\newcommand{\R}{\bldsy{\mathcal{R}}\lr{\bld{U}^{h}}}
\newcommand{\RV}{\bldsy{\mathcal{R}}^{\mom}\lr{\bld{U}^{h}}}
\newcommand{\RI}{\bldsy{\mathcal{R}}^{\ind}\lr{\bld{U}^{h}}}

\newcommand{\Ahmom}{\mathcal{A}^{\mom}\lr{\bld{W}^{h}, \bld{U}^{h}}}
\newcommand{\Ahind}{\mathcal{A}^{\ind}\lr{\bld{W}^{h}, \bld{U}^{h}}}
\newcommand{\M}{\mathcal{M}\lr{\bld{W}^{h}, \bld{U}^{h}; \bld{c},  h}}

\newcommand{\curl}{\nabla\times}
\renewcommand{\div}{\nabla\cdot}
\newcommand{\grads}{\nabla^{\mathrm{s}}}
\newcommand{\grada}{\nabla^{\mathsf{a}}}

\makeatletter
\newenvironment{tablehere}
  {\def\@captype{table}}
  {}

\makeatother

\begin{document}

\begin{frontmatter}
\title{A new class of finite element variational multiscale turbulence models for incompressible magnetohydrodynamics}
\author[uwm]{D.~Sondak\corref{cor1}}
\ead{sondak@math.wisc.edu}
\author[sandia]{J.N.~Shadid}
\author[rpi]{A.A.~Oberai}
\author[sandia]{R.P.~Pawlowski}
\author[sandia]{E.C.~Cyr}
\author[sandia]{T.M.~Smith}

\address[uwm]{University of Wisconsin-Madison, Department of Mathematcis}
\address[sandia]{Sandia National Laboratories, Computational Mathematics Department}
\address[rpi]{Rensselaer Polytechnic Institute, Department of Mechanical, Aerospace, and Nuclear Engineering}

\cortext[cor1]{Corresponding author}

\journal{J. Comput. Phys.}

\begin{abstract}
\input{abstract}
\end{abstract}


\end{frontmatter}

\input{Introduction}

\input{Background}
\input{TurbulenceModels}
\input{SolverDetails}
\input{Results}
\input{Conclusions}
\input{Acknowledgements}

\newpage

\bibliographystyle{plainnat}
\bibliography{mybib.bib}

\end{document}

%% file: abstract.tex
New large eddy simulation (LES) turbulence models for incompressible magnetohydrodynamics (MHD) derived from the variational multiscale (VMS) formulation for finite element simulations are introduced.  The new models include the variational multiscale formulation, a residual-based eddy viscosity model, and a mixed model that combines both of these component models.  Each model contains terms that are proportional to the residual of the incompressible MHD equations and is therefore numerically consistent.  Moreover, each model is also dynamic, in that its effect vanishes when this residual is small. The new models are tested on the decaying MHD Taylor Green vortex at low and high Reynolds numbers.  The evaluation of the models is based on comparisons with available data from direct numerical simulations (DNS) of the time evolution of energies as well as energy spectra at various discrete times.  A numerical study, on a sequence of  meshes, is 
presented that demonstrates that the large eddy simulation  approaches the DNS solution for these quantities with spatial mesh refinement.

%% file: Introduction.tex
\section{Introduction}
The study of fluid turbulence has a long and storied history that still lacks an explanation of the fundamental mechanisms driving this phenomenon~\cite{batchelor1982theory,pope2000turbulent}.  Despite this, the significance and ubiquity of turbulence drives us in trying to unlock its secrets.  An additional layer of complexity is added when considering electrically conducting fluids such as liquid metals and plasmas and their interaction with a magnetic field.   These types of fluids are even more common in the universe than non-electrically conducting fluids.  They are found in stars, at the cores of planets and in the interstellar medium.  On earth, engineers try to exploit their properties to generate electricity and fusion energy.  In most situations, these fluids are in a state of turbulence.  In an important subset of these cases, liquid metals, and under certain conditions plasmas, can be described by incompressible magnetohydrodynamics (MHD)~\cite{davidson2001introduction,goedbloed2006principles,goedbloed2010advanced}. 

The computational simulation of  turbulence is extremely challenging due to its inherent multiscale nature.  A perfect numerical experiment (direct numerical simulation, DNS) would capture all motions down to the point at which energy is dissipated as heat by molecular dissipation.  This is entirely infeasible for most problems and is likely to remain so for the foreseeable future.  The concept of large eddy simulation (LES) was introduced as a way of numerically resolving the largest-scales in the flow field while mathematically modeling the effect of the small-scales on the large-scales; see~\cite{rogallo1984numerical} and~\cite{sagaut2002large} for an overview of LES for hydrodynamics.
Several approaches for LES in MHD have been proposed.  In~\cite{theobald1994subgrid,haugen2006hydrodynamic,yoshizawa1991subgrid,muller2002large,ponty2004simulation,knaepen2004large,baerenzung2008spectrala,holm2002lagrangian} many techniques are presented including eddy viscosity models (EVMs), mean-field closures, and the Lagrangian-averaged MHD equations.  A comprehensive review of LES models in astrophysics can be found in~\cite{schmidt2014large}.

Most simulations of MHD turbulence rely on spectral numerical methods (see~\cite{peyret2010spectral} for background on spectral methods).  Although very fast (exponential convergence) and relatively 
straightforward to implement, these methods are not particularly amenable to the simulation of flows in complex geometries.  
The finite element (FE) method is an attractive alternative that enables computational simulation in both simple and complex geometries.  In the context of  fluid mechanics, and incompressible fluids in particular, there are a number of important issues that FE methods must address.
These include satisfaction of the incompressibility constraint, the inf-sup condition pertaining to conditions under which equal order elements for different fields may be used ~\cite{hughes1986new,hughes1989new}, and the control of spatial oscillations due to highly convected flow and unresolved internal and boundary layers.  Among the most popular and widely implemented solutions to these problems have been stabilized finite element methods~\citep{tezduyar1991stabilized, franca1992stabilized, codina2002stabilized, codina2007time}.  Around a decade after they were introduced, these methods were shown to derive from the general framework of the variational multiscale (VMS) method ~\cite{hughes1995multiscale, hughes1998variational}.  The VMS method was subsequently used to develop LES turbulence models for hydrodynamics~\cite{hughes2000large,bazilevs2007variational}.  

In spite of this work, LES models for MHD in the context of finite elements are relatively unexplored.  
Recent work has focused on developing accurate and robust stabilized finite element methods for incompressible MHD~\cite{gerbeau2000stabilized,codina2006stabilized,shadid2010towards,badia2013unconditionally,shadidetalMHD_2013}.  We refer to classical stabilized methods as incomplete turbulence models since they represent only part of the story coming from the VMS method.  In the context of spectral formulations VMS methods were used to derive new LES models for incompressible MHD in~\cite{sondak2012large}.  However, the development of complete VMS based LES models for MHD that can be implemented in a finite element context has not yet been explored.  This is the central theme of this manuscript.  In particular, the present work pursues two generalizations to previous work on a finite element based VMS model for the large eddy simulation of incompressible MHD flow:  (1)  the generalization of the spectral VMS turbulence models for MHD that were developed in~\cite{sondak2012large} and (2) the generalization of the VMS based finite element methods developed for incompressible MHD in \cite{shadidetalMHD_2013}.

The previous finite element based VMS models for MHD discussed above neglected second order correlations between unresolved scales (which we collectively refer to as ``Reynolds'' stresses) and included an incomplete version of the cross stresses.  The present work fully accounts for the effects of the cross and Reynolds stresses in MHD through the full VMS model and a VMS-based eddy viscosity model.  Our new models contain several new terms which include two cross stress terms in the momentum equation that are not present in classical stabilized formulations, two cross stress terms in the induction equation that are not present in classical stabilized formulations, Reynolds stresses in both the momentum and induction equations, and two VMS-based eddy viscosity models for the Reynolds stresses in the momentum and induction equations.

The layout of the remainder of the paper is as follows.  Section~\ref{sec:background} introduces the governing equations both in the strong form and variational forms.  Following this, section~\ref{sec:sim_details} describes the finite element method and the general computational framework.  Next, in section~\ref{sec:models}, a class of turbulence models is introduced for use with the finite element method.  A summary of numerical solution methods is presented in section~\ref{sec:solution_methods}. Section~\ref{sec:results} describes the test problem that was used to assess the model performance and discusses results.  Finally, section~\ref{sec:conclusions} reflects on the new models and considers improvements and other directions for the future.

%% file: Background.tex
\section{Background and Numerical Method}\label{sec:background}

\input{GoverningEquations}

\input{VariationalStatement}

\input{NumericalMethod}

%% file: GoverningEquations.tex
\subsection{Governing Equations}
The base model considered in this study is the incompressible resistive MHD system that is solved for a velocity field $\ub\lr{\bld{x},t} = \left[u_{1}, \ u_{2}, \ u_{3}\right]^{T}$ and a magnetic induction $\Bb\lr{\bld{x},t} = \left[B_{1}, \ B_{2}, \ B_{3}\right]^{T}$ for $\bld{x} = \left[x_{1}, \ x_{2}, \ x_{3}\right]^{T}$ in a domain $\Omega$ with boundary $\partial \Omega$.  We consider an incompressible velocity field so that $\nabla\cdot\ub = 0$.  The momentum equation is
\begin{align}
  \rho\pderiv{\ub}{t} + \rho\nabla\cdot\lr{\ub\otimes\ub} = \nabla\cdot\bld{T} + \bld{J}\times\Bb + \bld{f}^{\mom} \label{eq:mom}.
\end{align}
In the momentum equation, \eqref{eq:mom}, $\rho$ is the fluid density, $\bld{T}$ is the fluid stress tensor, $\bld{J}\times\Bb$  is  the Lorentz force where $\bld{J}$ is the current density,  and $\bld{f}^{V}$ is a general body force.  From the low frequency Maxwell's equations (specifically, Amp\`{e}re's law) we have $\bld{J} = \curl\Bb/\mu_{0}$ where $\mu_{0}$ is the magnetic permeability of free space.  This allows the Lorentz force to be rewritten as
\begin{align}
  \bld{J}\times\bld{B} = \frac{1}{\mu_{0}}\div\lr{\Bb\otimes\Bb} - \nabla\lr{\frac{1}{2\mu_{0}}\|\Bb\|^{2}}. \label{eq:FL}
\end{align}
Under the assumption of a Newtonian fluid, the stress tensor becomes
\begin{align}
  \bld{T} = -P\bld{I} + \bldsy{\Pi} \label{eq:T}
\end{align}
where $P$ is the fluid pressure, $\bld{I}$ is the identity tensor, and the viscous part of the stress tensor is given by
\begin{align}
  \bldsy{\Pi} = -\frac{2}{3}\mu\lr{\div\ub}\bld{I} + \mu\lr{\nabla\ub + \lr{\nabla\ub}^{T}}. \label{eq:Pi}
\end{align}
Using~\eqref{eq:FL},~\eqref{eq:T}, and~\eqref{eq:Pi} along with the divergence free constraint on the velocity field yields the final form of the momentum equation,
\begin{equation}
  \begin{split}
    &\rho\pderiv{\ub}{t} + \rho\div\lr{\ub\otimes\ub} - \frac{1}{\mu_{0}}\div\lr{\Bb\otimes\Bb} = \\
    &\hspace{2.0em}\nabla\cdot\lr{\mu\lr{\nabla\ub + \lr{\nabla\ub}^{T}}} - \nabla P - \nabla\lr{\frac{1}{2\mu_{0}}\|\Bb\|^{2}} + \bld{f}^{V}. \label{eq:mom_fin}
  \end{split}
\end{equation}

An equation for the magnetic induction is obtained by combining the low frequency Maxwell's equations (i.e. without a correction for the displacement current) with Ohm's law.  This leads to
\begin{align}
  \pderiv{\Bb}{t} + \div\lr{\ub\otimes\Bb - \Bb\otimes\ub} = \div\lr{\lambda\lr{\nabla\Bb - \lr{\nabla\Bb}^{T}}}
\end{align}
where $\lambda$ is the magnetic diffusivity.  The satisfaction of the solenoidal constraint on the magnetic induction, $\div\Bb=0$, has garnered considerable attention in the numerical community~\cite{toth2000b, codina2002stabilized}.  We take the approach of introducing a scalar Lagrange multiplier $\psi$ into the induction equation.  This permits the enforcement of the solendoidal condition as a divergence free constraint on the magnetic field.  Such an approach is common in finite element~\cite{codina2002stabilized, codina2011approximation} and finite volume methods~\cite{toth2000b, dedner2002hyperbolic, chacon2004non}.  Note that if a Fourier spectral method is used then the solenoidal constraints are satisfied identically and projections of $\psi$ into the Fourier spectral space are also identically zero.  Hence, in a pure Fourier spectral method the introduction of the scalar Lagrange multiplier is not necessary~\cite{sondak2012large}.  The final form of the induction equation is
\begin{align}
  \pderiv{\Bb}{t} + \div\lr{\ub\otimes\Bb - \Bb\otimes\ub} = \div\lr{\lambda\lr{\nabla\Bb - \lr{\nabla\Bb}^{T}}} - \nabla \psi + \bld{f}^{\ind} \label{eq:ind_fin}
\end{align}
where we have also included $\bld{f}^{\ind}$ as a possible external electromagnetic force.

We introduce some notation to simplify the subsequent exposition.  First, denote the vector of solutions by $\bld{U} = \left[\ub, \ P, \ \Bb, \ \psi \right]^{T}$.  Then the nonlinear terms in the momentum equation are given by
\begin{align}
  \nlv\lr{\bld{U}} = \rho\ub\otimes\ub - \frac{1}{\mu_{0}}\Bb\otimes\Bb + \frac{1}{2\mu_{0}}\|\Bb\|^{2}\bld{I}.
\end{align}
The viscous part of the stress tensor is
\begin{align}
  \bld{F}^{\mom}\lr{\bld{U}} = \mu\lr{\nabla\ub + \lr{\nabla\ub}^{T}}.
\end{align}
Similarly, the nonlinear terms in the induction equation are given by
\begin{align}
  \nli\lr{\bld{U}} = \ub\otimes\Bb - \Bb\otimes\ub
\end{align}
and the magnetic diffusive flux is
\begin{align}
  \bld{F}^{\ind}\lr{\bld{U}} = \lambda\lr{\nabla\Bb - \lr{\nabla\Bb}^{T}}.
\end{align}
Putting this notation together, the final form of the incompressible MHD equations is
\begin{align}
  &\rho\pderiv{\ub}{t} + \div\lr{\nlv\lr{\bld{U}} - \bld{F}^{\mom}\lr{\bld{U}}} + \nabla P = \bld{f}^{\mom} \label{eq:mom_final} \\
  &\pderiv{\Bb}{t} + \div\lr{\nli\lr{\bld{U}} - \bld{F}^{\ind}\lr{\bld{U}}} + \nabla \psi = \bld{f}^{\ind} \label{eq:ind_final} \\
  &\hspace{3.0em}\div\ub = \div\Bb = 0 \label{eq:sol_constraints}.
\end{align}

%% file: VariationalStatement.tex
\subsection{Variational Statement of the Governing Equations}
We use a finite element method to solve the MHD system (\eqref{eq:mom_final},~\eqref{eq:ind_final}, and~\eqref{eq:sol_constraints}) in a domain $\Omega$ with either periodic boundary conditions for all variables or homogeneous Dirichlet boundary conditions for the velocity and the induction fields.  Both of these boundary conditions imply that the space of trial and weighting functions are the same.  We note that we have made this choice in order to simplify the notation.  The methods described in this paper can easily be extended to more complex boundary conditions.  Before discussing the numerical approach, we begin by stating the variational counterpart to the system described above:  \textit{Find $\bld{U} \in \V$ s.t. $\forall \  \bld{W} \in \V$}
\begin{align}
     \mathcal{A}\lr{\bld{W}, \bld{U}} = \lr{\bld{W}, \bld{F}} \label{eq:abstract_var}
\end{align}
where $\bld{W} = \left[\bld{w}, \ q, \bld{v}, \ s\right]^{T}$ is a vector of weighting functions and $\bld{F} = \left[\bld{f}^{\mom}, \ 0, \bld{f}^{\ind}, \ 0\right]^{T} $ is a vector of forcing functions.  The notation $\lr{\cdot, \cdot}$ represents an $L_{2}$ inner product of two functions,
\begin{align}
     \lr{\bld{u}, \bld{v}} = \int{\bld{u}\cdot\bld{v} \ \mathrm{d}\Omega}.
\end{align}
We must specify the function space $\V$ in which the solutions and weighting functions reside.  This function space consists of functions that are sufficiently smooth and satisfy the requisite boundary conditions and is given by
\begin{equation}
  \begin{split}
     &\V \equiv \{\bld{W} \ | \ \bld{W} = \left[\bld{w}, \ q, \ \bld{v}, \ s\right]^{T} \ \mathrm{;} \\
 &\hspace{2.5em} w_{i}, \ v_{i} \in H^{1}\lr{\Omega}, q, \ s \in L_{2}\lr{\Omega} \}. \label{eq:function_space}
  \end{split}
\end{equation}
In~\eqref{eq:function_space}, $L_{2}$ denotes the space of square-integrable functions while $H^{1}$ is the space of square-integrable functions with square-integrable derivatives.  The semilinear form in~\eqref{eq:abstract_var} is given by
\begin{align}
     \mathcal{A}\lr{\bld{W},\bld{U}} &= \mathcal{A}^{\mom}\lr{\bld{W}, \bld{U}} - \lr{\nabla q, \ub} + \mathcal{A}^{\ind}\lr{\bld{W}, \bld{U}} - \lr{\nabla s, \Bb} \label{eq:A} \\
     \lr{\bld{W}, \bld{F}}           &= \lr{\bld{w}, \bld{f}^{\mom}} + \lr{\bld{v}, \bld{f}^{\ind}} \label{eq:WF}
\end{align}
where
\begin{align}
     \mathcal{A}^{\mom}\lr{\bld{W}, \bld{U}} &= \lr{\bld{w}, \rho\pderiv{\ub}{t}} -
          \lr{\nabla\bld{w}, \nlv\lr{\bld{U}}} - \lr{\div\bld{w}, P} + 
          \lr{\nabla\bld{w},\bld{F}^{\mom}\lr{\bld{U}}} \label{eq:AVform}    \\
     \mathcal{A}^{\ind}\lr{\bld{W}, \bld{U}} &= \lr{\bld{v}, \pderiv{\Bb}{t}} -
          \lr{\nabla\bld{v}, \nli\lr{\bld{U}}} + \lr{\bld{v}, \nabla \psi} + 
          \lr{\nabla\bld{v}, \bld{F}^{\ind}\lr{\bld{U}}}. \label{eq:AIform}
\end{align}
Note that no boundary terms appear in~\eqref{eq:AVform} and~\eqref{eq:AIform} as a result of the boundary conditions of the problem.

%% file: NumericalMethod.tex
\subsection{Finite Element Discretization}\label{sec:sim_details}

We are now prepared to introduce the finite element method.  The numerical solution consists of selecting specific functions from the function space $\V$ that provide reasonable approximations to the actual functions that represent the solutions to the equations.  That is $\bld{U}^{h} \approx \bld{U}$.  Hence, we will select functions from a conforming, finite-dimensional subspace of $\V$ which we denote $\V^{h}$.   Mathematically, $\V^{h} \subset \V$.  The parameter $h$ indicates the finite dimensional space or a function that was chosen from the space $\V^{h}$.  In practical terms, $h$ represents the mesh parameter and determines the size of the finite elements being employed.

We denote an element in physical space as $K$ and an element in parametric space as $K^{e}$.  Coordinates in physical space are $\bld{x} = {x_{i}}, \ i=1,2,3$ and coordinates in parametric space are $\bldsy{\xi} = {\xi_{i}}, \ i= 1, 2, 3$.  We consider an isoparamteric mapping,
\begin{align}
  \bld{x}\lr{\bldsy{\xi}} = \sum_{a=1}^{8}\bld{x}^{e}_{a}\phi_{a}\lr{\bldsy{\xi}}
\end{align}
where $\phi_{a}\lr{\bldsy{\xi}}$ are trilinear hexahedral basis functions and $\bld{x}^{e}_{a}$ are the coordinates of the eight element nodes.  The solution is given by
\begin{align}
  \bld{U}^{h}\lr{\bldsy{\xi}} = \sum_{a=1}^{8}\bld{U}^{e}_{a}\phi_{a}\lr{\bldsy{\xi}} \label{eq:Uea}
\end{align}
and the discretized weighting functions are given by
\begin{align}
  \bld{W}^{h}\lr{\bldsy{\xi}} = \sum_{a=1}^{8}\bld{W}^{e}_{a}\phi_{a}\lr{\bldsy{\xi}}. \label{eq:Wea}
\end{align}
The Galerkin statement is therefore:  \textit{Find $\bld{U}^{h} = \left[\ub^{h}, \ P^{h}, \ \Bb^{h}, \ \psi^{h}\right]^{T} \in \V^h$ s.t. $\forall \ \bld{W}^{h}=\left[\bld{w}^{h}, \ q^{h}, \ \bld{v}^{h}, \ s^{h}\right]\in\V^{h}$}
\begin{align}
     \mathcal{A}\lr{\bld{W}^{h}, \bld{U}^{h}} = \lr{\bld{W}^{h}, \bld{F}^{h}} \label{eq:abstract_var_h}
\end{align}
where $\mathcal{A}\lr{\cdot,\cdot}$ is given by~\eqref{eq:A}.

A very serious pitfall of the Galerkin approximation is that the numerical solution is very rarely able to capture all of the details of the flow field.  This is particularly true for turbulent flows wherein multiple scales are active simultaneously.  Very often, however, one is interested only in relatively large scales of the flow field.  Considering the large scales in isolation is entirely inadequate and, at the very least, the effect of the small scales on the large scales must be taken into account.  This is the principle behind large eddy simulation; that is, mathematical models (called turbulence models) are developed that model the effects of the small scales on the large scales.  This is the topic of the next section.  

%% file: TurbulenceModels.tex
\section{Turbulence Models}\label{sec:models}
     Several new turbulence models were developed and implemented in~\citep{sondak2012large}.  The models were implemented in a Fourier-spectral setting.  There are several differences in the form that the models take depending on the numerical method used.  In particular, the models require more terms when using non-orthogonal basis functions as is done in the finite element setting.  The first set of models that we present are derived from the variatonal multiscale formulation under the assumption of scale separation.

\input{VMS}

\input{EddyViscosityModels}

\input{MixedModel}

\input{GeneralModelExpression}

%% file: VMS.tex
\subsection{Variational Multiscale Formulation}
       With the assumption of scale separation, we will model the effects below a certain length scale and numerically resolve the motion above that length scale.  Mathematically we say that the motions are represented by the finite dimensional functions $\bld{U}^{h}$.  To be more precise, we say that the solutions $\bld{U}^h$ are optimal when they are defined via a projection operator $\mathbb{P}^{h}$.  That is,
\begin{align}
     \bld{U}^{h} = \mathbb{P}^{h}\bld{U} \in \V^{h}.
\end{align}
The projection operator $\mathbb{P}^{h}$ projects the analytical solution onto the finite dimensional subspace in an optimal manner.  The user determines the meaning of optimal.  Examples include ensuring that the finite dimensional space contains functions that are nodally exact or functions that minimize some type of error norm.  The choice of the projection operator $\mathbb{P}^{h}$ induces a fine-scale projection operator $\mathbb{P}^{'}$ which allows the introduction of the fine scale solution,
\begin{align}
     \bld{U}^{'} = \mathbb{P}^{'}\bld{U} \in \V^{'}.
\end{align}
Thereafter we introduce a decomposition of the solution field into coarse-scales and fine-scales,
\begin{align}
     \bld{U} = \bld{U}^{h} + \bld{U}^{'} \label{eq:u_decomp}.
\end{align}
This procedure is the starting point of the variational multiscale formulation as introduced in~\citep{hughes1995multiscale} and extended to large eddy simulation in~\citep{bazilevs2007variational}.  The decomposition in~\eqref{eq:u_decomp} is also used for the weighting functions so that
\begin{align}
     \bld{W} = \bld{W}^{h}+ \bld{W}^{'} \label{eq:w_decomp}.
\end{align}
These decompositions imply a function space decomposition of the form
\begin{align}
     \V = \V^{h} \oplus \V^{'}.
\end{align}
We note that the choice of projection operator will determine the nature of the function space decomposition.  In~\citep{sondak2012large}, the projection operators were selected to be the high-pass and low-pass sharp cutoff filters.  These projection operators result in orthogonal subspaces which in turn possess convenient properties that can be exploited.  For example, inner products between functions in $\V^{h}$ and $\V^{'}$ are zero due to the orthogonality property.  In this work, we leave the projection operators unspecified.  A careful consideration of projection operators giving rise to orthogonal subspaces in a finite element context for fluid mechanics is presented in~\citep{codina2002stabilized}.  A thorough discussion of projection operators is also considered in~\citep{hughes2008variational}.

Introducing the decompositions~\eqref{eq:u_decomp} and~\eqref{eq:w_decomp} into~\eqref{eq:abstract_var} results in two coupled problems,
\begin{align}
     &\mathcal{A}\lr{\bld{W}^{h},\bld{U}^{h} + \bld{U}^{'}} = \lr{\bld{W}^{h}, \bld{F}}, \quad \forall \ \bld{W}^{h} \in \V^{h} \label{eq:coarse}      \\
     &\mathcal{A}\lr{\bld{W}^{'},\bld{U}^{h} + \bld{U}^{'}} = \lr{\bld{W}^{'}, \bld{F}}, \quad \forall \ \bld{W}^{'} \in \V^{'} \label{eq:fine}   .
\end{align}
The goal is to solve the first of these problems for $\bld{U}^{h}$ and the second for $\bld{U}^{'}$.  Up to this point these two problems are exact.  The variational multiscale formulation proceeds by finding an approximate solution to~\eqref{eq:fine}.  The approximate solution to~\eqref{eq:fine} is (see~\citep{bazilevs2007variational} for the hydrodynamic case,~\citep{badia2013unconditionally} and references therein for MHD),
\begin{align}
     \bld{U}^{'} \approx -\bldsy{\tau}\R\label{eq:uprime}.
\end{align}
In~\eqref{eq:uprime}, $\R$ is the residual of the coarse scales and is given by
\begin{align}
     \R =
          \begin{bmatrix}
               \RV               \\
               \div\ub^{h} \\
               \RI                \\
               \div\Bb^{h}
          \end{bmatrix}
\end{align}
where
\begin{align}
     \RV = \rho\pderiv{\ub^{h}}{t} + \div\lr{\nlv\lr{\bld{U}^{h}} - \bld{F}^{\mom}\lr{\bld{U}^{h}}} + \nabla P^{h} - \bld{f}^{\mom}
\end{align}
is the momentum residual and
\begin{align}
     \RI = \pderiv{\Bb^{h}}{t} + \div\lr{\nli\lr{\bld{U}^{h}} - \bld{F}^{\ind}\lr{\bld{U}^{h}}} + \nabla \psi^{h}  - \bld{f}^{\ind} 
\end{align}
is the induction residual.  The parameter $\bldsy{\tau}$ is the stabilization matrix and is an algebraic approximation to the inverse differential operator on the fine scales.  Moreover, the stabilization matrix is an intrinsic grid time-scale consisting of a combination of advective and diffusive time-scales.  In this work, it is represented as a diagonal matrix operator $\bldsy{\tau} = \textrm{diag}\lr{\tau^{\mom}, \tau^{\mom}, \tau^{\mom}, \tau^{\mom}_{c}, \tau^{\ind}, \tau^{\ind}, \tau^{\ind}, \tau^{\ind}_{c}}$.  The diagonal components are given by
\begin{align}
     \tau^{\mom} &= 
          \frac{1}{\sqrt{\lr{\frac{2C_{t}^{\mom}\rho}{\Delta t}}^{2} +
          \rho^{2}\ub^{h}\cdot\bld{G}\ub^{h} + 
          \lr{C_{\mu}\mu}^{2}\|\bld{G}\|^{2} + 
          \frac{C_{B}^{2}\rho}{\mu_{0}}\|\bld{B}^{h}\|^{2}\|\bld{G}\|}}
          \label{eq:tau_mom}
          \\
     \tau^{\ind} &=
          \frac{1}{\sqrt{\lr{\frac{2C_{t}^{\ind}}{\Delta t}}^{2} +
          \ub^{h}\cdot\bld{G}\ub^{h} + 
          \bld{B}^{h}\cdot\bld{G}\bld{B}^{h} +
          C_{\lambda}\lambda^{2}\|\bld{G}\|^{2}     }    }
          \label{eq:tau_ind}
          \\
     \tau^{\mom}_{c} &= \frac{1}{C_{t}^{\mom}\textrm{tr}\lr{\bld{G}}\tau^{\mom}} 
          \label{eq:tau_c}
          \\
     \tau^{\ind}_{c}    &= \frac{1}{C_{t}^{\ind}\textrm{tr}\lr{\bld{G}}\tau^{\ind}}
          \label{eq:tau_r}
\end{align}
where the components of $\bld{G}$ are given by
\begin{align}
     G_{ij} = \pderiv{\xi_{k}}{x_{i}}\pderiv{\xi_{k}}{x_{j}}
\end{align}
and
\begin{align}
     \mathrm{tr}\lr{\bld{G}} = G_{ii}. \quad \text{(Summation on $i$ implied)}
\end{align}
The constants in these expressions are $\mathcal{O}\lr{1}$ and do not change upon mesh refinement or coarsening.  Ultimately, this fact does have implications for the influence of the stabilization parameter on the solution.  The constants were selected to be $C_{t}^{\mom} = 1$, $C_{t}^{\ind} = 1$, $C_{\mu} = 1$, $C_{B} = 1$, and $C_{\lambda} = 1$.  The VMS statement is:  \textit{Find $\bld{U}^{h} \in \V^{h}$ s.t. $\forall \ \bld{W}^{h} \in \V^{h}$ }
\begin{equation}
  \begin{split}
     &\Ahmom - \lr{\nabla\bld{w}^{h}, \nlv_{C}\lr{\bld{U}^{h}, \bld{U}^{\prime}} + \nlv_{R}\lr{\bld{U}^{\prime}}} - \lr{\div\bld{w}^{h}, P^{\prime}}            \\
     &\hspace{3em} + \lr{\nabla\bld{w}^{h}, \bld{F}^{\prime \ \mom}\lr{\bld{U}^{\prime}}} - \lr{\nabla q^{h}, \ub'}     \\
     +&\Ahind    - \lr{\nabla \bld{v}^{h}, \nli_{C}\lr{\bld{U}^{h}, \bld{U}^{\prime}} + \nli_{R}\lr{\bld{U}^{\prime}}} + \lr{\bld{v}^{h}, \nabla \psi'}                                            \\
     &\hspace{3em} + \lr{\nabla\bld{v}^{h}, \bld{F}^{\prime \ \ind}\lr{\bld{U}^{\prime}}} - \lr{\nabla s^{h}, \Bb'}    \\
     &=\lr{\bld{w}^{h}, \bld{f}^{\mom}} + \lr{\bld{v}^{h}, \bld{f}^{\ind}} . \label{eq:vms_full}
  \end{split}
\end{equation}
In~\eqref{eq:vms_full} the cross stresses for the momentum and induction equations are respectively given by
\begin{align}
     \nlv_{C}\lr{\bld{U}^{h}, \bld{U}^{\prime}} &=  \rho\lr{\ub^{h}\otimes\ub' + \ub'\otimes\ub^{h}} - \frac{1}{\mu_{0}}\lr{\Bb^{h}\otimes\Bb' + \Bb'\otimes\Bb^{h}} + \frac{1}{\mu_{0}}\Bb^{h}\cdot\Bb' \label{eq:cross_mom} \\
     \nli_{C}\lr{\bld{U}^{h}, \bld{U}^{\prime}} &= \lr{\ub^{h}\otimes\Bb' - \Bb'\otimes\ub^{h}} + \lr{\ub'\otimes\Bb^{h}-\Bb^{h}\otimes\ub'} \label{eq:cross_ind}
\end{align}
while the Reynolds stress contributions to the momentum and induction equations are given by
\begin{align}
     \nlv_{R}\lr{\bld{U}^{\prime}} &= \rho\lr{\ub'\otimes\ub'} - \frac{1}{\mu_{0}}\lr{\Bb'\otimes\Bb'} + \frac{1}{2\mu_{0}}\|\Bb'\|^{2} \label{eq:reynolds_stresses}     \\
     \nli_{R}\lr{\bld{U}^{\prime}} &= \lr{\ub'\otimes\Bb'} - \lr{\Bb'\otimes\ub'} \label{eq:induction_stresses} .
\end{align}
In~\eqref{eq:vms_full} $\bld{F}^{\prime \ \mom}$ and $\bld{F}^{\prime \ \ind}$ are the viscous stress tensors arising from the subgrid scales where
\begin{align}
  \bld{F}^{\prime \ \mom}\lr{\bld{U}^{\prime}} &= \mu\lr{\nabla\ub' + \lr{\nabla\ub'}^{T}}       \\
  \bld{F}^{\prime \ \ind}\lr{\bld{U}^{\prime}} &= \lambda\lr{\nabla\Bb' - \lr{\nabla\Bb'}^{T}}.
\end{align}
It was demonstrated in~\citep{wang2010spectral} in the context of hydrodynamic turbulence that using the first order approximation to $\bld{U}^{'}$,~\eqref{eq:uprime}, results in a negligible contribution from the Reynolds stresses~\eqref{eq:reynolds_stresses}.  Following this example, we neglect the second order correlations given by~\eqref{eq:reynolds_stresses} and~\eqref{eq:induction_stresses}.  Note that to date there is no precedent for dropping the term $\dfrac{1}{2\mu_{0}}\|\Bb'\|^{2}$ in~\eqref{eq:reynolds_stresses}.  We ran simulations both with and without this term and found no discernable effects on the results and therefore concluded that its contribution is negligible. Using linear finite elements the inner products that involve gradients of the subgrid scales can be shown to be zero through an additional integration by parts.  The final finite element VMS statement is: \textit{Find $\bld{U}^{h}\in\V^{h}$ s.t. $\forall \ \bld{W}^{h} \in \V^{h}$},
\begin{equation}
   \begin{split}
     &\Ahmom - \lr{\nabla\bld{w}^{h}, \nlv_{C}\lr{\bld{U}^{h}, \bld{U}^{\prime}}} - \lr{\div\bld{w}^{h}, P^{\prime}} - \lr{\nabla q^{h}, \ub'} \\
     +&\Ahind    - \lr{\nabla \bld{v}^{h}, \nli_{C}\lr{\bld{U}^{h}, \bld{U}^{\prime}}} + \lr{\bld{v}^{h}, \nabla \psi'} - \lr{\nabla s^{h}, \Bb'} \label{eq:vms_implemented} \\
     &=\lr{\bld{w}^{h}, \bld{f}^{\mom}} + \lr{\bld{v}^{h}, \bld{f}^{\ind}}. 
   \end{split}
\end{equation}
If the fine scale space was designed to be truly orthogonal to the finite element space then all inner products between functions in the fine scale and coarse scale spaces would vanish identically.  This is the situation that is found when using a Fourier-spectral method as in~\citep{sondak2012large}.  For a detailed, general discussion of this property the reader is referred to~\citep{codina2002stabilized} and references therein.

%% file: EddyViscosityModels.tex
\subsection{Eddy Viscosity Models}
We briefly review some developments in eddy viscosity models for MHD.  The general form of the models presented in this work include any of the eddy viscosity models reviewed in this section.  In particular, the eddy viscosity models reviewed here are described in~\cite{theobald1994subgrid},~\cite{muller2002large}, and~\citep{sondak2012large}.

\subsubsection{Classic Eddy Viscosity Models}
In~\citep{sondak2012large}, the VMS-based turbulence modeling approach was compared to the classical Smagorinsky LES model for MHD developed in~\citep{theobald1994subgrid} as well as an extension to this classic model that takes into account fundamental MHD physics~\citep{muller2002large}.  The variational form of the MHD equations inclusive of an eddy viscosity model is: \textit{Find $\bld{U}^{h}\in\V^{h}$ s.t. $\forall \ \bld{W}^{h} \in \V^{h}$}
\begin{align}
     \mathcal{A}\lr{\bld{W}^{h},\bld{U}^{h}} +  \M = \lr{\bld{W}^{h}, \bld{F}} \label{eq:var_model}
\end{align}
where the model term $\mathcal{M}\lr{\cdot,\cdot; \cdot, \cdot}$ seeks to model the effects of the fine scales on the coarse scales.  This model term generally depends on the coarse-scale solution, some model parameters $\bld{c}$, and the mesh parameter $h$.  The classical Smagorinsky eddy viscosity model for MHD was introduced in~\citep{theobald1994subgrid}.  In a variational context the model term takes the form,
\begin{align}
     \M = \lr{\nabla\bld{w}^{h}, 2\rho\nu_{T}\grads\ub^{h}} + \lr{\nabla\bld{v}^{h}, 2\lambda_{T}\grada\Bb^{h}} \label{eq:mhd_evm}
\end{align}
where
\begin{align*}
  \grads\ub^{h} &= \frac{1}{2}\lr{\nabla\ub^{h} + \lr{\nabla\ub^{h}}^{T}} \\
  \grada\Bb^{h} &= \frac{1}{2}\lr{\nabla\Bb^{h} - \lr{\nabla\Bb^{h}}^{T}}.
\end{align*}
The turbulent viscosity $\nu_{T}$ and turbulent magnetic diffusivity $\lambda_{T}$ are given by
\begin{align}
     \nu_{T}     &= C_{\mom}^{S}{h}^{2}\mathcal{S}, \quad \mathcal{S} = \sqrt{2\grads\ub^{h}\bldsy{:}\grads\ub^{h}} \label{eq:nut_dsev}     \\
     \lambda_{T} &= C_{\ind}^{S}{h}^{2}\left|\sqrt{\frac{\mu_{0}}{\rho}}\bld{J}^{h}\right|, \quad \bld{J}^{h} = \frac{1}{\mu_{0}}\curl\Bb^{h} \label{eq:lambdat_dsev}.
\end{align}
When the coefficients $C_{\mom}^{S}$ and $C_{\ind}^{s}$ are determined via a dynamic procedure~\citep{germano1991dynamic, oberai2005dynamic} the models are referred to as the dynamic Smagorinsky eddy viscosity (DSEV) model.  

An extension to the model introduced in~\citep{theobald1994subgrid} was introduced in~\citep{muller2002large}.  We refer to this extension as the alignment-based dynamic Smagorinsky eddy viscosity (DSEVA) model.  The model has the same form as~\eqref{eq:mhd_evm} with the only difference being in the definition of the turbulent diffusivities.  These are now given by
\begin{align}
     \nu_{T}             &= C_{\mom}^{S}h^{2}\sqrt{\textrm{abs}\lr{\grads\ub^{h}\bldsy{:}\frac{1}{\mu_{0}\rho}\grads\Bb^{h}}} \label{eq:nut_aligned} \\
     \lambda_{T}     &=  C_{\ind}^{S}h^{2}\textrm{sgn}\lr{\bld{J}^{h}\cdot\bldsy{\omega}^{h}} 
                                             \sqrt{\textrm{abs}\lr{\sqrt{\frac{\mu_{0}}{\rho}}\bld{J}^{h}\cdot\bldsy{\omega}^{h}}} \label{eq:lambdat_aligned}.
\end{align}
In~\eqref{eq:nut_aligned} and~\eqref{eq:lambdat_aligned}, $\textrm{sgn}\lr{\cdot}$ gives the sign of its argument, $\textrm{abs}\lr{\cdot}$ gives the absolute value of its argument, and $\bldsy{\omega}^{h}$ is the fluid vorticity.  The coefficients are still determined via the classic dynamic procedure.  The DSEVA model takes into account the alignment of the velocity field and magnetic induction which is a fundamental characteristic of MHD flows.

\subsubsection{Residual-Based Eddy Viscosity Model}
A new type of eddy viscosity model was proposed in~\citep{oberai2014residual, sondak2012large}.  This eddy viscosity model was motivated via the VMS formulation and has the same form as~\eqref{eq:var_model} with the model in the form of~\eqref{eq:mhd_evm}.  The eddy diffusivities are given by
\begin{align}
     \nu_{T} = \lambda_{T} = \overline{C}h\sqrt{\left|\ub'\right|^{2} + \frac{1}{\mu_{0}\rho}\left|\Bb'\right|^{2}} \label{eq:rbev_diff}
\end{align}
where $\overline{C}$ is a universal constant.  The constant is given by
\begin{align}
  \overline{C} = \frac{2}{3\sqrt{3}C_{K}^{3/2}hk^{h}\sqrt{1-\beta^{-2/3}}} \label{eq:Cbar}
\end{align}
where
$h$ is the mesh size, $C_{K} = 2.2$ is the dimensionless Kolmogorov constant (see~\cite{beresnyak2011spectral, biskamp2003magnetohydrodynamic}), $\beta$ is a constant denoting the range of the numerical subgrid solutions, and $k^{h}$ is the cutoff wavenumber.  This expression is derived from a more general expression in~\cite{sondak2012large} which was determined by matching the numerical and exact dissipation rates.  In~\eqref{eq:Cbar} we have assumed that the energy spectrum exhibits a $k^{-5/3}$ power law behavior in the inertial range where $k$ is the wavenumber.  We note that the true inertial range behavior in MHD is still undetermined and that numerical evidence suggests a nonuniversal character for the MHD energy spectrum~\cite{lee2010lack}.  Recent theoretical studies have indicated that the spectrum in MHD turbulence is actually $k^{-3/2}$~\cite{boldyrev2006spectrum}.  However, a $k^{-5/3}$ energy spectrum leads to a universal expression for $\overline{C}$ that does not depend on the dissipation rate.  In light of this, we choose $\overline{C}$ as given by~\eqref{eq:Cbar} and acknowledge that modificiations to the derivation of $\overline{C}$ may be appropriate in the future.

In a spectral setting the cutoff wavenumber is easily determined to be $k^{h} = \pi/h$.  In a finite element setting we choose $k^{h} = \pi/\lr{2h}$ by reasoning that it requires four linear elements to resolve the smallest wavelength in the domain.  Moreover, we choose $\beta=3/2$ but note that $\beta < 3/2$ may be more appropriate when using linear finite elements.  By choosing $\beta = 3/2$ we are assuming that the subgrid solutions have wavelengths between $k^{h}$ and $3k^{h}/2$.  With these parameters and assumptions, we obtain $\overline{C} \approx 0.15$.

In~\eqref{eq:rbev_diff} both $\ub'$ and $\Bb'$ are given by~\eqref{eq:uprime} and are therefore based off of the residual of the coarse scales.  Hence this model is automatically dynamic and it is conjectured that the constant $\overline{C}$ does not need to be adjusted dynamically.  Because the expressions for the eddy diffusivities involve formulae for the subgrid scales derived from the VMS formulation we refer to this eddy viscosity model as the residual-based eddy viscosity (RBEV) model.

%% file: MixedModel.tex
\subsection{Mixed Model}
For large values of fluid and magnetic Reynolds numbers, $Re$ and $Rm$, the second-order correlations ($\nlv_{R}\lr{\bld{U}^{\prime}}$ and $\nli_{R}\lr{\bld{U}^{\prime}}$) are important and we would expect that the pure VMS models given by~\eqref{eq:vms_implemented} do not sufficiently account for the subgrid scales.  Motivated by this, a mixed model for MHD was proposed in~\citep{sondak2012large}.  In the present formulation, this model takes the form,
\begin{equation}
  \begin{split}
     &\Ahmom - \lr{\nabla\bld{w}^{h}, \nlv_{C}\lr{\bld{U}^{h}, \bld{U}^{\prime}}} - \lr{\div\bld{w}^{h}, P'} - \lr{\nabla q^{h}, \ub'} \\
     &\hspace{3.0em} + \mathcal{M}_{\mom}\lr{\bld{W}^{h}, \bld{U}^{h}; \bld{c}^{\mom}, h} \\
   +&\Ahind    - \lr{\nabla \bld{v}^{h}, \nli_{C}\lr{\bld{U}^{h}, \bld{U}^{\prime}}} + \lr{\bld{v}^{h}, \nabla \psi'} - \lr{\nabla s^{h}, \Bb'} \\
     &\hspace{3.0em} + \mathcal{M}_{\ind}\lr{\bld{W}^{h}, \bld{U}^{h}; \bld{c}^{\ind}, h} \\
     &=\lr{\bld{w}^{h}, \bld{f}^{\mom}} + \lr{\bld{v}^{h}, \bld{f}^{\ind}} \label{eq:mixed_model}
  \end{split}
\end{equation}
where the additional model terms $\mathcal{M}_{\mom}\lr{\cdot, \cdot; \cdot, \cdot}$ and $\mathcal{M}_{\ind}\lr{\cdot, \cdot; \cdot, \cdot}$ are eddy viscosity models of the form~\eqref{eq:mhd_evm}.  The eddy viscosities in the models could be given by any of~\eqref{eq:nut_dsev}-\eqref{eq:lambdat_aligned} or~\eqref{eq:rbev_diff}.  When the mixed model uses the RBEV model to account for the second order correlations it is a fully residual-based model.

%% file: GeneralModelExpression.tex
\subsection{General Expression for Residual-Based Models}
For completeness, we present a general expression for the models.  Note that the formulation contains parameters that may be adjusted depending on the desired model.  Specific values of the parameters are provided in Table~\ref{tab:b_parameters} that result in the particular models developed.
\begin{equation}
  \begin{split}
     &\Ahmom - b_{\VMS}^{\mom}\left[\lr{\nabla\bld{w}^{h}, \nlv_{C}\lr{\bld{U}^{h}, \bld{U}^{\prime}}} + \lr{\div\bld{w}^{h}, P'}\right] - c_{\VMS}^{\mom}\lr{\nabla q^{h},\ub'} \\
     &\hspace{2.0em} + b_{\EVM}^{\mom}\lr{\nabla\bld{w}^{h}, 2\rho\nu_{T}\grads\ub^{h}} \\
   +&\Ahind - b_{\VMS}^{\ind}\left[\lr{\nabla \bld{v}^{h},\nli_{C}\lr{\bld{U}^{h}, \bld{U}^{\prime}}} - \lr{\bld{v}^{h}, \nabla \psi'}\right] - c_{\VMS}^{\ind}\lr{\nabla s^{h}, \Bb'} \label{eq:all_models}  \\
     &\hspace{2.0em} + b_{\EVM}^{\ind}\lr{\nabla\bld{v}^{h}, 2\lambda_{T}\grada\Bb^{h}} \\
   =& \lr{\bld{w}^{h}, \bld{f}^{\mom}} + \lr{\bld{v}^{h}, \bld{f}^{\ind}} .
  \end{split}
\end{equation}
Note that~\eqref{eq:all_models} includes the DSEV and DSEVA models as possible choices by selecting the eddy diffusivities according to~\eqref{eq:nut_dsev}-\eqref{eq:lambdat_dsev} or~\eqref{eq:nut_aligned}-\eqref{eq:lambdat_aligned} respectivly.  Moreover, a fully residual-based model can be attained by selecting the eddy diffusivities according to~\eqref{eq:rbev_diff}.
\tablehere{
\begin{table}
\caption{A summary of the values of the parameters $b_{i}$ and $c$ and corresponding models.}    
\centering
\begin{center}
\begin{tabular}{cccc}
\toprule
\multicolumn{1}{c}{Model} & \multicolumn{1}{c}{$b^{\mom, \ind}_{\textrm{VMS}}$} 
                          & \multicolumn{1}{c}{$b^{\mom, \ind}_{\textrm{EVM}}$}
                          & \multicolumn{1}{c}{$c^{\mom, \ind}_{\textrm{VMS}}$} \\
\midrule
VMS   & 1 &  0    & 1  \\ \midrule
MM    & 1 &  1    & 1  \\ \midrule
\multirow{2}*{EVM} & \multirow{2}*{0} & \multirow{2}*{1}      & 1 \ \text{equal order elements} \\
                                      &                       &                                   &0  \ \text{any other time \qquad \ } \\
\bottomrule
\end{tabular}
\end{center}
\label{tab:b_parameters}
\end{table}
}
Finally, the parameter $c_{\VMS}^{\mom,\ind}$ deserves comment.  This parameter is necessary depending on the type of finite element basis functions that are being used.  Classically, this term was introduced to circumvent the inf-sup condition in mixed problems.  That is, it allows equal order elements for all of the fields in the problem.  Note that this term comes for free when using VMS-based methods.  However, when using pure EVMs one needs to be careful and introduce the term only when appropriate as indicated in Table~\ref{tab:b_parameters}.  For more details on the inf-sup condition and the addition of the stabilization term the reader is referred to~\citep{hughes1986new}.

%% file: SolverDetails.tex
\section{Summary of Fully-implicit AMG Preconditioned Newton-Krylov Solution}\label{sec:solution_methods}

\subsection{Fully-implicit Time Integration}
The incompressible  restive MHD model, \eqref{eq:mom_final} - \eqref{eq:sol_constraints}, supports a number of fast time-scales.
These include elliptic effects due to the divergence constraints that support 
infinite speed pressure waves and enforce the solenoidal involution of the magnetic field, fast time-scale parabolic effects due to momentum and magnetic diffusion, 
and Alfv\'{e}n wave propagation. This system can exhibit behavior that is characterized by widely separated time-scales as well as overlapping time-scales.  These characteristics make the robust and efficient iterative solution of these systems very challenging.  In this context fully-implicit methods are an 
attractive choice that can often provide unconditionally-stable time 
integration techniques.  These methods can be designed with various types 
of stability properties that allow robust integration of multiple-time-scale 
systems without the requirement to resolve the stiff modes of the system 
(which are not of interest since they do not control the accuracy of time 
integration). In the computations presented in this study, a fully-implicit  
multistep backward differentiation method of order three (BDF3) is employed
to accurately and efficiently integrate the time evolution of the MHD system. More details on the fully-implicit integration and example order-of-accuracy 
verification results can be found in \cite{shadid2010towards,shadidetalMHD_2013}.

\subsection{Strongly-coupled Newton-Krylov Solver} 
The result of a fully-implicit solution technique is the development of very large-scale,
 coupled highly nonlinear algebraic  system(s) that must be solved. 
Therefore, these techniques place a heavy burden on both the nonlinear and 
linear solvers and require robust, scalable, and efficient iterative solution 
methods. In this study Newton-based iterative nonlinear solvers 
(\cite{Dennis-Schnabel}) are employed to solve the challenging nonlinear systems 
that result from this application. These solvers can exhibit quadratic convergence 
rates independently of the problem size when sufficiently robust linear solvers 
are available. For the latter, we employ Krylov iterative techniques. A 
Newton-Krylov (NK) method is an implementation of Newton's 
method in which a Krylov iterative solution technique is used to approximately 
solve the linear systems, $\bf{J}_k\bf{s}_{k+1}=-\bf{F}_k$, that are generated at 
each step of Newton's method. 
For efficiency, an inexact Newton method (\cite{Dennis-Schnabel}, ~\cite{EW96}) is 
usually employed, whereby one approximately solves the linear systems generated 
in the Newton method by choosing a forcing term  $\eta_k$ and stopping the Krylov 
iteration when the inexact Newton condition, 
$ \|\bf{F}_k+\bf{J}_k\bf{s}_{k+1}\|\leq\eta_{k+1}\|\bf{F}_k\| $ is satisfied. The particular 
Krylov method that is used in this study is a robust non-restarted GMRES method
that is capable of iteratively converging to the solution of very large non-symmetric linear systems 
such as developed from the discretized resistive MHD system provided a
sufficiently robust and scalable preconditioning method is available \cite{shadid2010towards,shadidetalMHD_2013}.

\subsection{Fully-coupled Algebraic Multilevel Preconditioner}
As a scalable preconditioner for the discretized VMS resistive MHD system 
a fully-coupled algebraic multigrid method (AMG) is employed. In general AMG methods
are significantly easier to implement and  integrate with complex unstructured mesh simulation software than
geometric multigrid methods~\cite{tuminaro-cnme,shadid2004,Shadid2006}.
AMG preconditioners associate a graph with the matrix 
system being solved. Graph vertices correspond to matrix rows for scalar PDEs, while for
PDE systems it is natural to associate one vertex with each nodal block of 
unknowns (e.g. velocities, pressure, magnetic field and the Lagrange multiplier at a particular grid point).
Note that  since the VMS formulation allows the use of non-mixed FE interpolation
(in this case for example all first order Lagrange interpolation on hexes) a simple block matrix
(8x8) structure exists if all unknowns are ordered consecutively at each FE vertex and a 
graph edge exists between vertex $i$ and $j$ if there is a 
nonzero in the block matrix which couples $i$'s rows with $j$'s columns 
or $j$'s rows with $i$'s columns. The specific  AMG coarsening strategy 
that is employed is a parallel greedy graph aggregation technique
which attempts to make an aggregate by taking an unaggregated point
and grouping it with all of its neighbors. Thus, it tends to coarsen
by a factor of three in each coordinate direction when applied to a 
standard discretization matrix obtained by a compact stencil on a regular 
mesh. In addition, a dynamic load-balancing package, Zoltan~\cite{devine2002zoltan},  is used to repartition coarsened operators 
across processors. This generally improves parallel performance and gives
better aggregates on the next coarser level which is obtained by again
applying the parallel aggregation technique. A more complete discussion of this strategy 
and the implementation for this method can be found in~\cite{shadidetalMHD_2013}.
The fully-coupled AMG preconditioner and the Newton-GMRES method described above 
have been demonstrated to provide a robust and scalable nonlinear/linear iterative solution strategy
for a number of applications that include MHD (see e.g. \cite{shadid2010towards,shadidetalMHD_2013,lin2010AMGjcp} and the references contained therein).

\subsection{Nonlinear and Linear Convergence Criteria}
To complete the brief description of the numerical method that is employed in this study 
the convergence criteria that are used at each time-step to determine convergence of the nonlinear/linear iterative methods must be stated.
For the nonlinear solver two convergence criteria are employed.  The first
is a requirement of sufficient reduction in the relative nonlinear residual norm, $\| F_{k} \| / \|
F_{o} \| < 10^{-2}$.   In general, this requirement is set to be easily satisfied.
The second convergence criterion is based on a sufficient decrease of
a weighted norm of the relative size of the Newton update vector.  The latter
criterion requires that the correction, $\Delta \chi _{i}^{k}$, for any
grid point unknown, $\chi _{i}$, is small compared to its magnitude, $\left|\chi
_{i}^{k}\right|$, and is given by
\[
\sqrt{\frac{1}{N_{u}}\sum _{i=1}^{N_{u}}\left[\frac{\left|\Delta \chi _{i}\right|}{\varepsilon _{r}\left|\chi _{i}\right|+\varepsilon _{a}}\right]^{2}}<1,
\]
where $N_{u}$ is the total number of unknowns, $\varepsilon _{r}$ is
the relative error tolerance between the variable correction and its
magnitude, and $\varepsilon _{a}$ is the absolute error tolerance of
the variable correction, which
sets the magnitude of components that are to be considered
to be numerically zero.  In this study, the relative-error and
absolute-error tolerance are $10^{-3}$ and $10^{-6}$ respectively.
In practice this criteria controls convergence of the 
nonlinear solver to sufficient accuracy. For example, in the high Reynolds number $256^3$ VMS simulation (see section~\ref{sec:highReresults})
the maximum relative nonlinear norm decrease was $\mathcal{O}(10^{-5})$.
Finally in each Newton sub-step the linear system is solved to a relative tolerance of
 $\eta = 10^{-3}$ with the knowledge that the overall accuracy of the solution for each time step 
 is controlled as described above for convergence of the Newton step.

%% file: Results.tex
\section{Results and Discussion}\label{sec:results}

The models are evaluated on the Taylor-Green vortex generalized to MHD as described in~\citep{pouquet2010dynamics} and~\citep{lee2010lack}.  This problem begins with smooth initial conditions for the velocity field and magnetic induction,
\begin{align}
     \mathbf{u}\lr{\bld{x},t=0} = u_{0}
		\begin{bmatrix}
			 \sin\lr{x}\cos\lr{y}\cos\lr{z} \\
			-\cos\lr{x}\sin\lr{y}\cos\lr{z} \\
			 0
		\end{bmatrix}	.
\end{align}
\begin{align}
\mathbf{B}\lr{\bld{x},t=0} = B_{0}
		\begin{bmatrix}
			  \cos\lr{x}\sin\lr{y}\sin\lr{z} \\
			  \sin\lr{x}\cos\lr{y}\sin\lr{z} \\
			-2\sin\lr{x}\sin\lr{y}\cos\lr{z}
		\end{bmatrix}    .
\end{align}
The domain is a periodic box of size $[-\pi,\pi]^{3}$.  The energy of each field is determined from their respective energy spectrum,
\begin{align}
         K^{\tot,\mom,\ind} = \int{E^{\tot,\mom,\ind}\lr{k} \ \text{d}k}
\end{align}
      where $K^{\tot,\mom,\ind}$ represents the total energy, kinetic energy, or magnetic induction energy at a certain time and  $E^{\tot, \mom,\ind}\lr{k}$ is the total energy spectrum, the kinetic energy spectrum, or the magnetic energy spectrum at wavenumber $k$.  We define the Reynolds number $Re$ and magnetic Reynolds number $Rm$ as
\begin{align}
         Re &= \frac{u_{\text{rms}}L^{\mom}}{\nu} \\
         Rm &= \frac{u_{\text{rms}}L^{\ind}}{\lambda}.
\end{align}
   The characteristic length scales $L^{\mom}$ and $L^{\ind}$ are computed as
\begin{align}
         L^{\mom,\ind} = \frac{\int\frac{1}{k}E^{\mom,\ind}\lr{k}\text{d}k}{\int E^{\mom,\ind}\lr{k}\text{d}k}.
\end{align}  
      The ratio of molecular viscosity to magnetic diffusivity, the magnetic Prandtl number ($Pr=Rm/Re$), is unity.  The initial velocity and magnetic fields are scaled so that the total energy, which is the sum of the kinetic and magnetic energies, is initially equal to $0.25$.  The initial state also has the kinetic and magnetic energies equal to each other. For details on properties of this flow field, the reader is referred to~\citep{lee2010lack, pouquet2010dynamics, dallas2013origins} and references therein.  In particular,~\citep{lee2010lack} discusses the nonuniversality of MHD by considering multiple Taylor-Green initial conditions and demonstrating that these different initial conditions lead to different energy cascades.  In~\citep{dallas2013origins}, the energy spectrum of the particular Taylor-Green vortex considered in this work is discussed as well as some implications of nonuniversality for turbulence modeling in MHD.  However, we also note the interesting conclusions in~\cite{aluie2010scale} on locality of MHD turbulence which is an essential ingredient for a universal behvior of the energy cascade.
      
\input{LowReResults}
\input{HighReResults}

%% file: LowReResults.tex
\subsection{Low Reynolds Number}

\begin{figure}[H]
  	\centering
    \includegraphics[width=110mm,height=70mm]{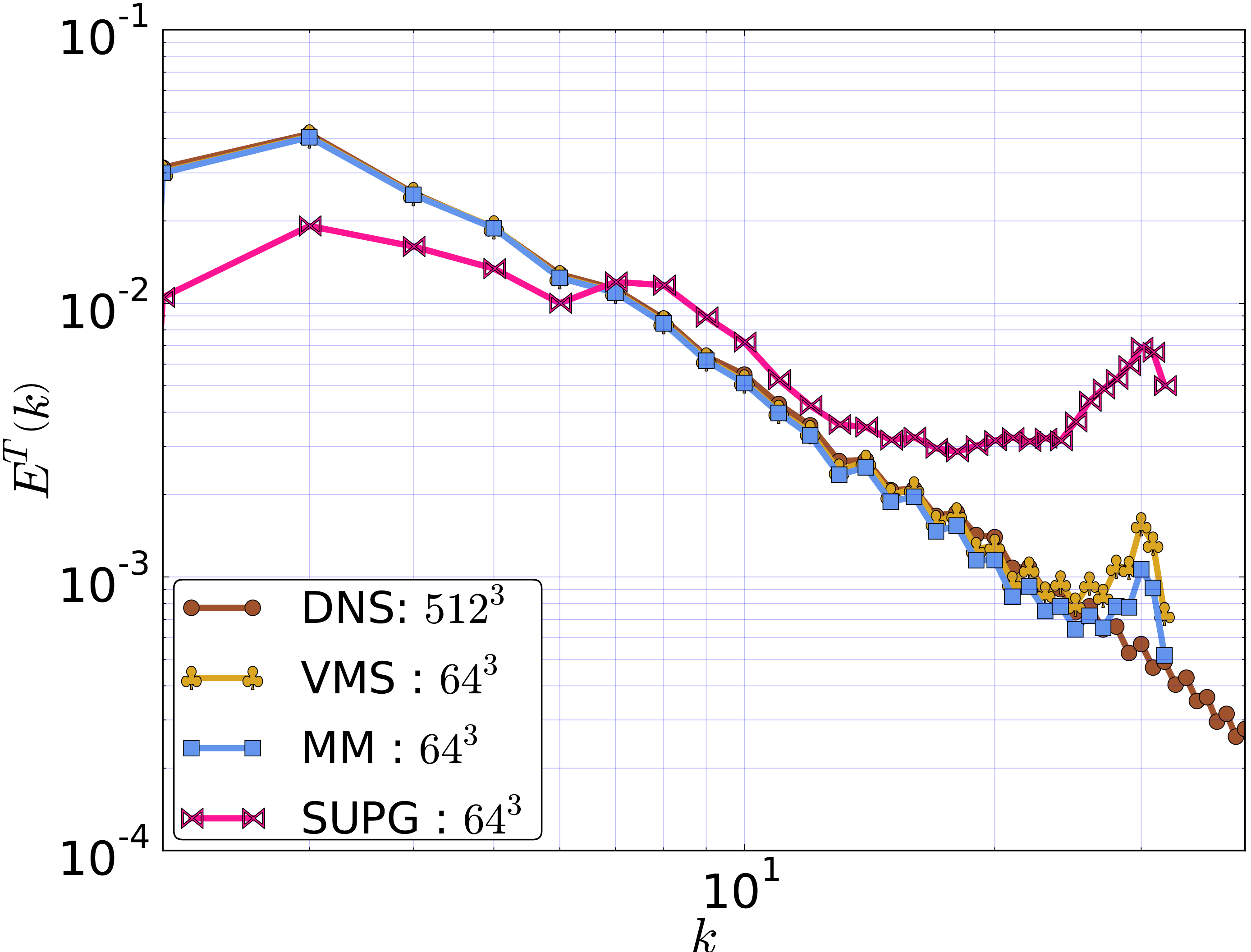}
    \caption{The total energy spectrum for the low Re case. Here the SUPG stabilized method is compared with the VMS model, the mixed model, and to a DNS simulation.}
    \label{fig:SUPG4}
  \end{figure}

We first consider the case of $Re=Rm=1800$ at the peak of dissipation.  For this problem, the peak of dissipation occurs at $t\in[3.75, \ 4.25]$.  To motivate the necessity of more sophisticated models in MHD, we first consider the classical SUPG model adapted to MHD.  This model can be obtained as a partial VMS decomposition.  That is, we decompose the velocity field in the momentum equation and only keep the upwinding term.  Similarly, the VMS decomposition is used on the magnetic induction in the induction equation and we once again only keep the upwinding term.  Critically, the SUPG formulation neglects energy transfer from the subgrid magnetic induction to the momentum equation and from the subgrid velocity field to the induction equation.  Mathematically, the cross stress terms in equations~\eqref{eq:cross_mom} and~\eqref{eq:cross_ind} are reduced to
\begin{align*}
     \nlv_{\text{SUPG}}\lr{\bld{U}^{h}, \bld{U}^{\prime}} &=  \rho\ub'\otimes\ub^{h} \\
     \nli_{\text{SUPG}}\lr{\bld{U}^{h}, \bld{U}^{\prime}} &= \lr{\ub^{h}\otimes\Bb' - \Bb'\otimes\ub^{h}}
\end{align*}
while the Reynolds stresses are neglected entirely.  Figure~\ref{fig:SUPG4} compares the total energy spectrum at $t=4$ from the SUPG stabilization to the VMS and mixed models.  All energy spectra have been averaged over neighboring modes to help remove the nodal oscillations that are typically present in energy spectra obtained from the Taylor Green vortex.  From these results it is clear that more than a straightforward SUPG stabilization is required to sufficiently dissipate the unresolved highest wavenumber modes.  Even the low wavenumber modes are not accurately represented.  This can be attributed to the fact that we are plotting the total energy and that the SUPG formulation does not account for cross coupling of subgrid fields.  We note that with mesh refinement the SUPG results would improve because the effects of the subgrid fields would be diminished.  Indeed, even with the VMS and mixed models there is still a slight pile-up of energy.  Note, however, that the mixed model is able to dissipate the energy in the last modes despite the slight energy pile-up.  This is the desired behavior although we will have further comments on the energy pile up in subsequent sections.

\begin{figure}[H]
  \centering
  \includegraphics[width=110mm]{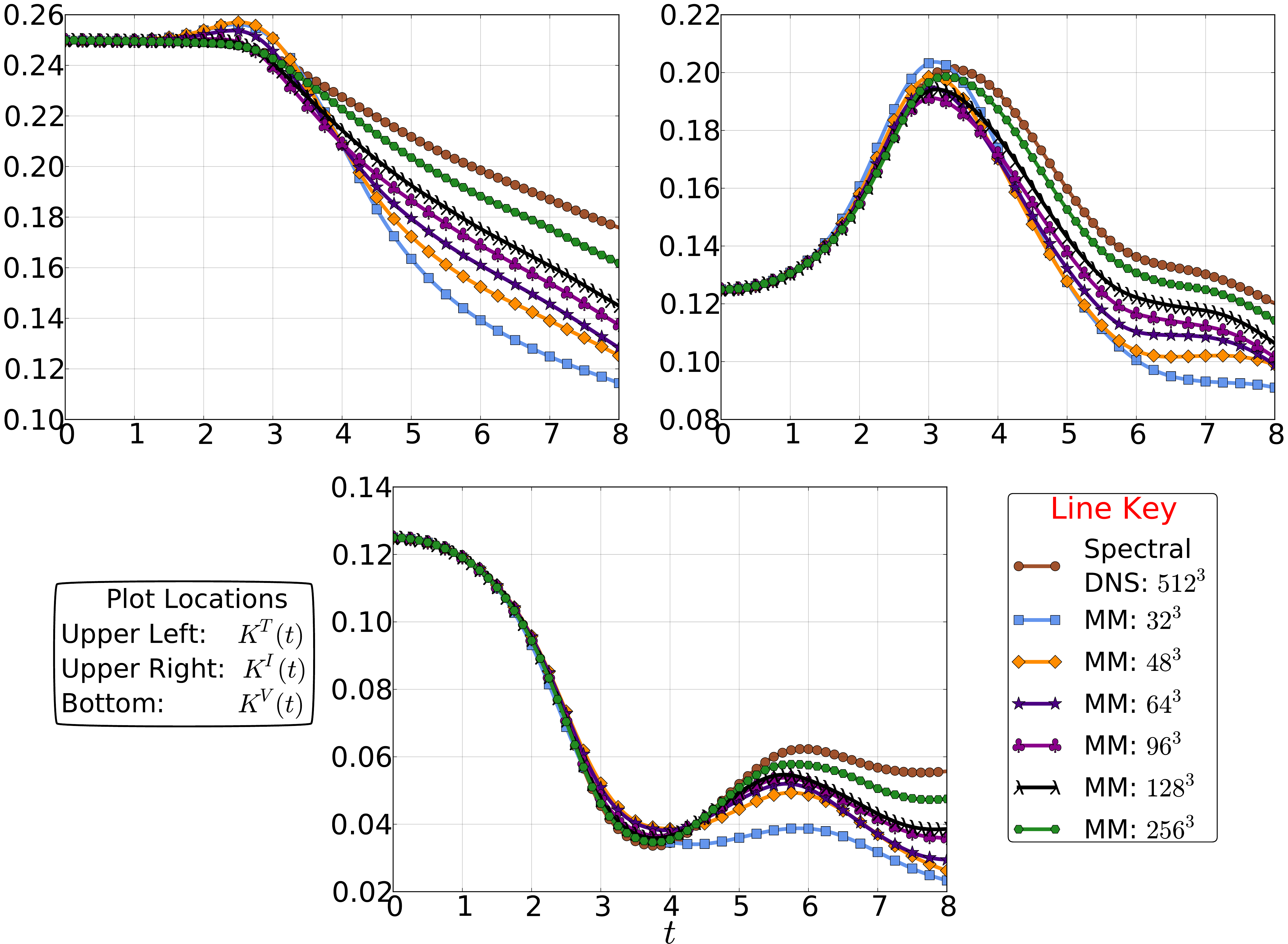}
  \caption{Time evolution of MHD energies from the mixed model (MM) LES MHD turbulence model at various mesh resolutions.}
  \label{fig:Ke_mc}
\end{figure}

\begin{figure}[H]
    \centering
    \includegraphics[width=110mm,height=60mm]{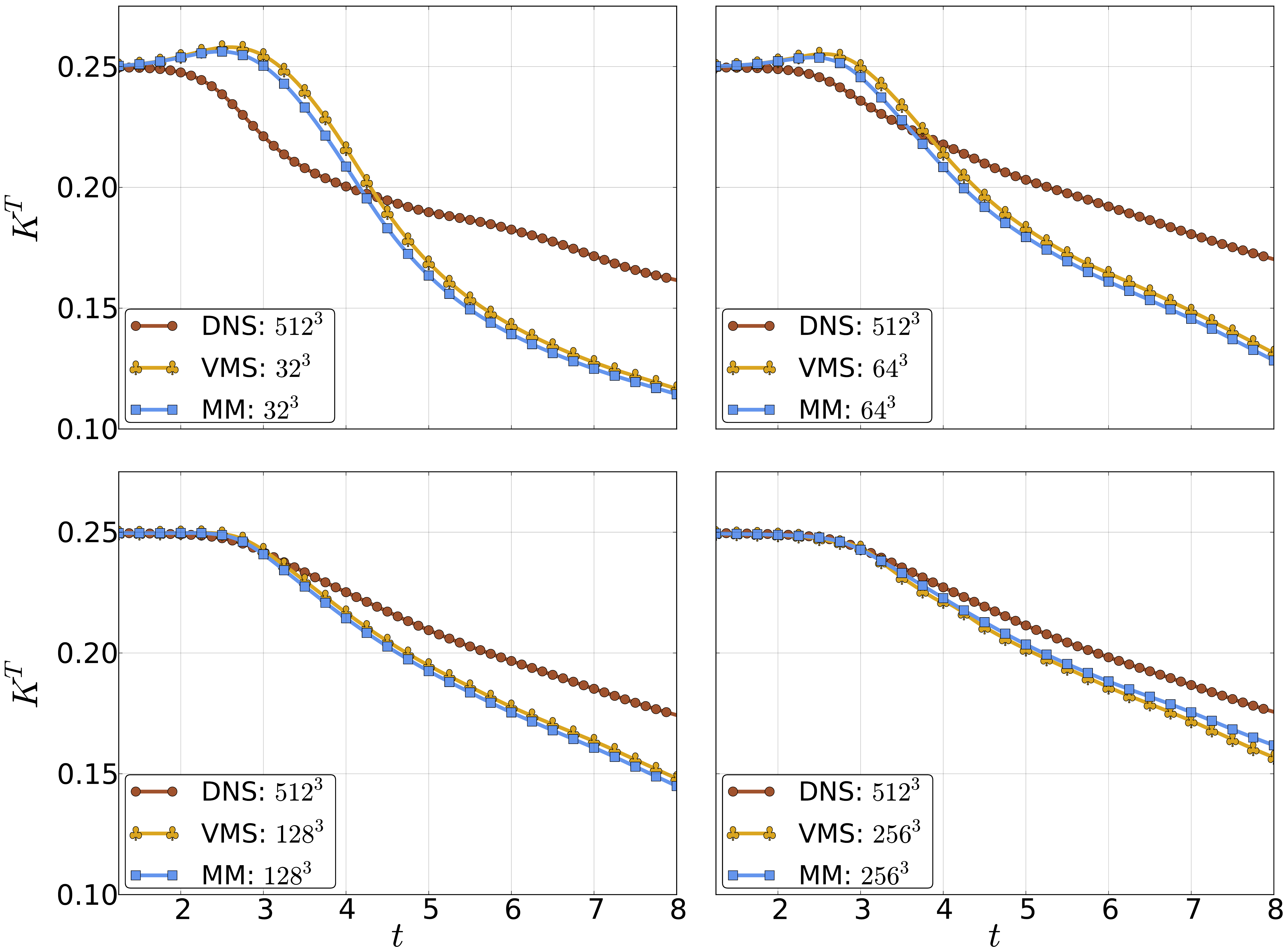}
    \caption{Time evolution of total energy.  The DNS results have been low-pass filtered such that the number of Fourier modes matches the number of finite elements in the LES model.}
    \label{fig:KeT}
  \end{figure}

\begin{figure}[H]
    \centering
    \includegraphics[width=110mm,height=60mm]{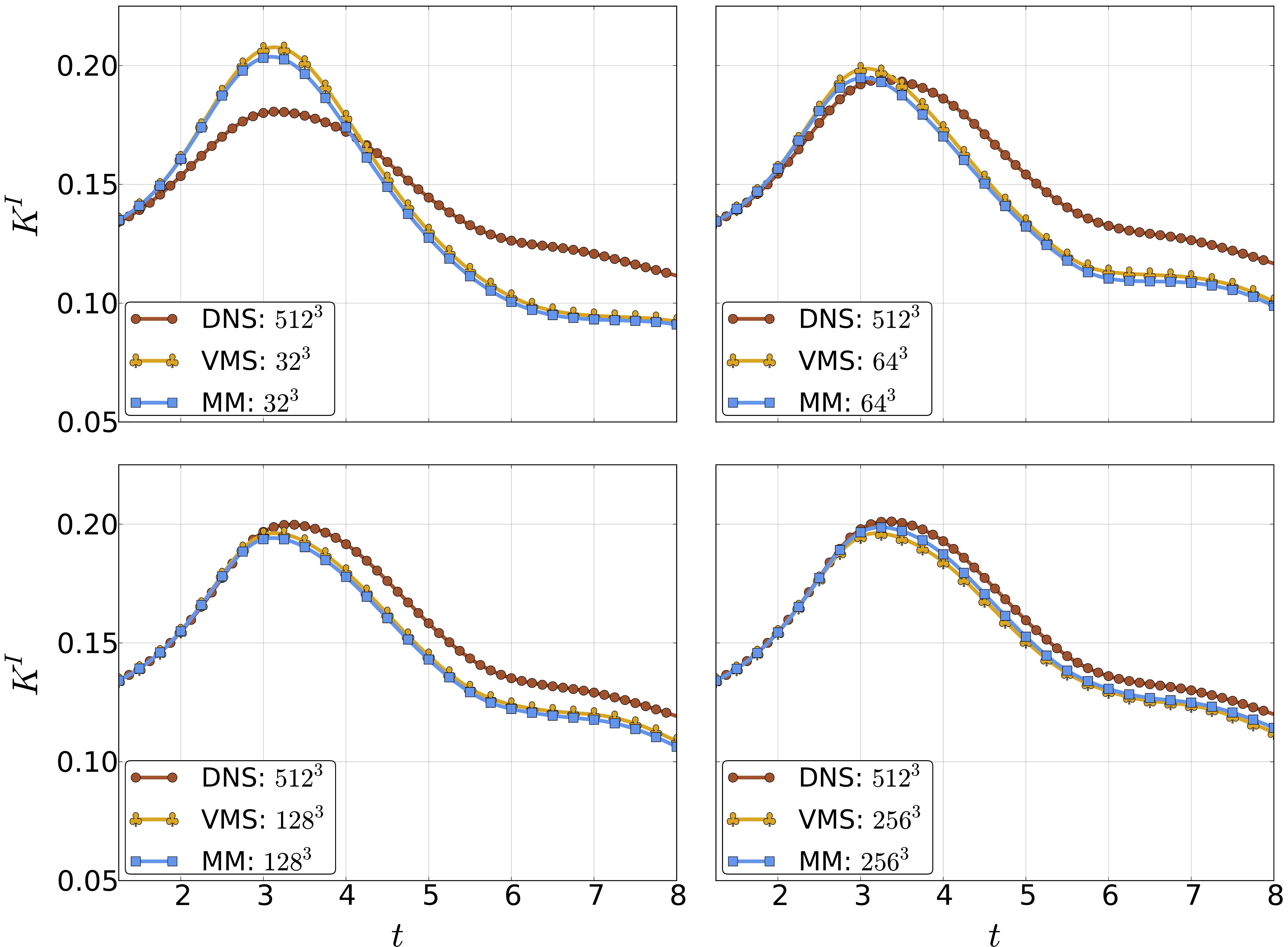}
    \caption{Time evolution of magnetic energy.  The DNS results have been low-pass filtered such that the number of Fourier modes matches the number of finite elements in the LES model.}
    \label{fig:KeI}
  \end{figure}

\begin{figure}[H]
    \centering
    \includegraphics[width=110mm,height=60mm]{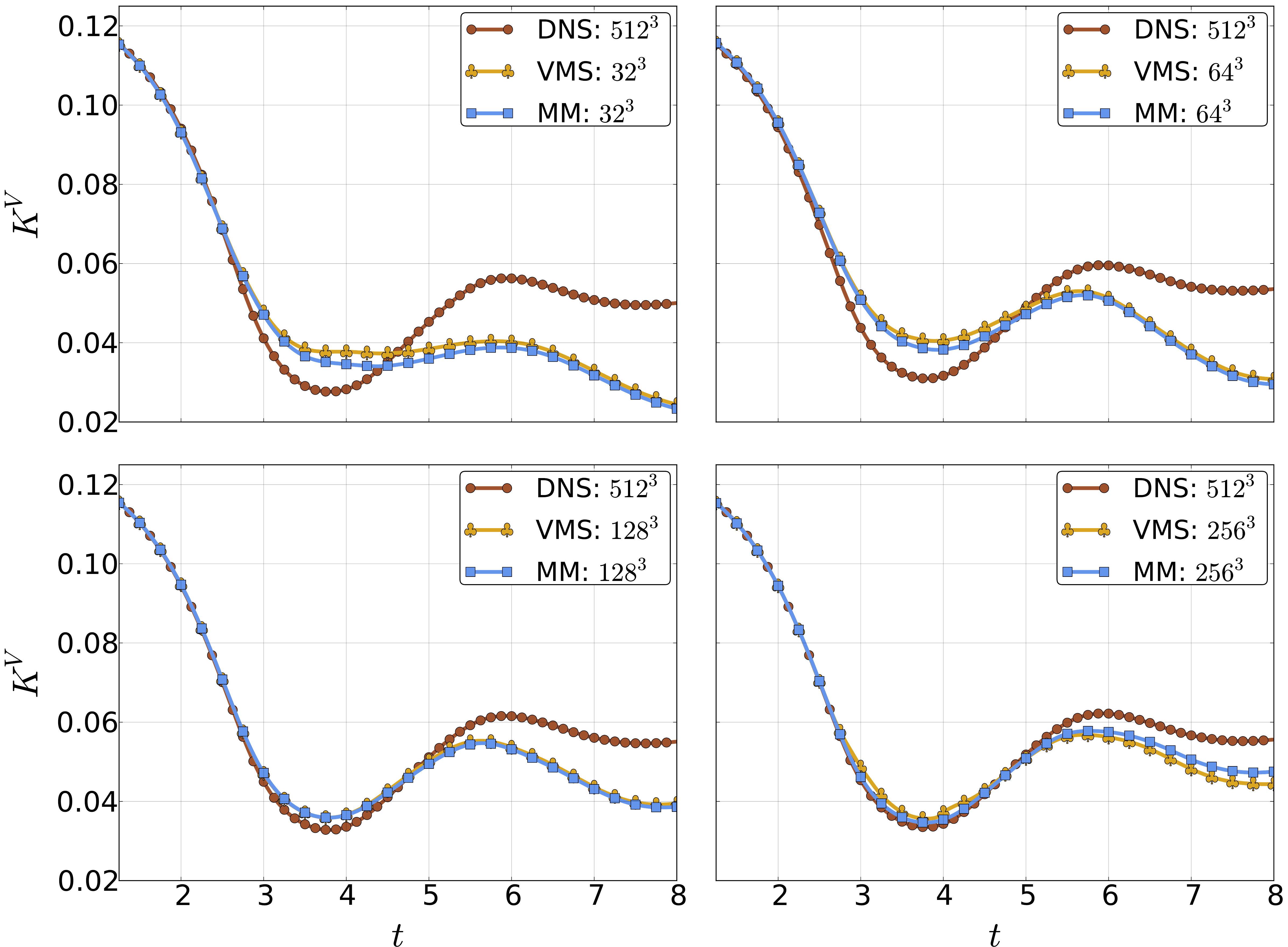}
    \caption{Time evolution of kinetic energy.  The DNS results have been low-pass filtered such that the number of Fourier modes matches the number of finite elements in the LES model.}
    \label{fig:KeV}
  \end{figure}

In Figure~\ref{fig:Ke_mc} we consider the time evolution of MHD energies for various LES mesh resolutions ranging from $N=32$ to $N=256$.    The LES results are compared to a spectral DNS at $N=512$.  It is observed that the models qualitatively converge monotonically with mesh refinement to the DNS results.
The efficacy of the new LES models is further analyzed in Figures~\ref{fig:KeT}-\ref{fig:KeV}.  Once again, the time evolution of MHD energies computed from the VMS model and mixed model compared to the spectral DNS is presented.  The turbulence models are presented at four different resolutions: $N=32$, $N=64$, $N=128$, and $N=256$.  The DNS results have been low-pass filtered in order to compare with the FE LES solutions.  This however is not a straightforward  comparison since the FE simulations clearly have a much lower convergence rate (i.e. formally second-order) than the spectral method that is exponentially convergent. It does however provide some qualitative assessment of the relative convergence of this class of models for this problem compared to the spectral DNS approximation.  We first note that the solutions improve considerably with mesh refinement.

Indeed, at $N=64$ both models are in quite good agreement for times up to about $t =4$, the peak of dissipation, with the mixed model performing marginally better.  We observe, however, that both models have a tendency to overpredict the total energy through the peak of dissipation for coarse discretizations.  This is a result of the VMS approach and the fact that we did not ensure the orthogonality of the subgrid scales with the resolved scales.  In the spectral setting one can prove that the VMS method is globally dissipative~\citep{wang2010spectral} (although local backscatter is possible).  Such a demonstration is facilitated by the orthogonality of two spaces involved.  We note that for $N=128$ and beyond the over-prediction near the peak of dissipation is no longer present and the results correspond to the DNS solution quite well.

\begin{figure}[H]
    \centering
    \includegraphics[width=110mm, height=60mm]{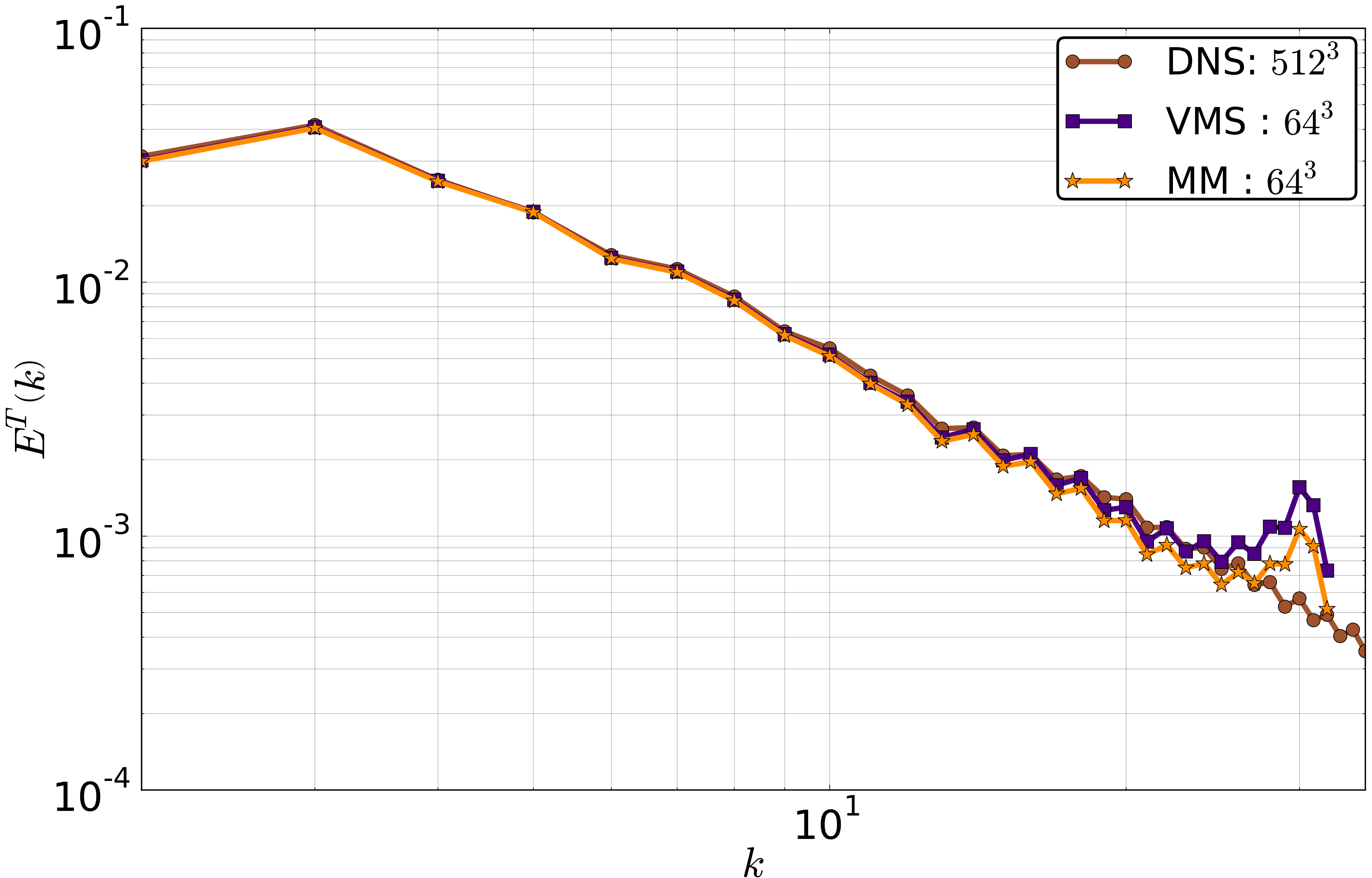}
    \caption{The total energy spectrum for the VMS model and mixed model compared to the DNS simulation at $t=4$.}
    \label{fig:VMSvsMM}
\end{figure}

\begin{figure}[H]
  	  \centering
      \includegraphics[width=110mm,height=60mm]{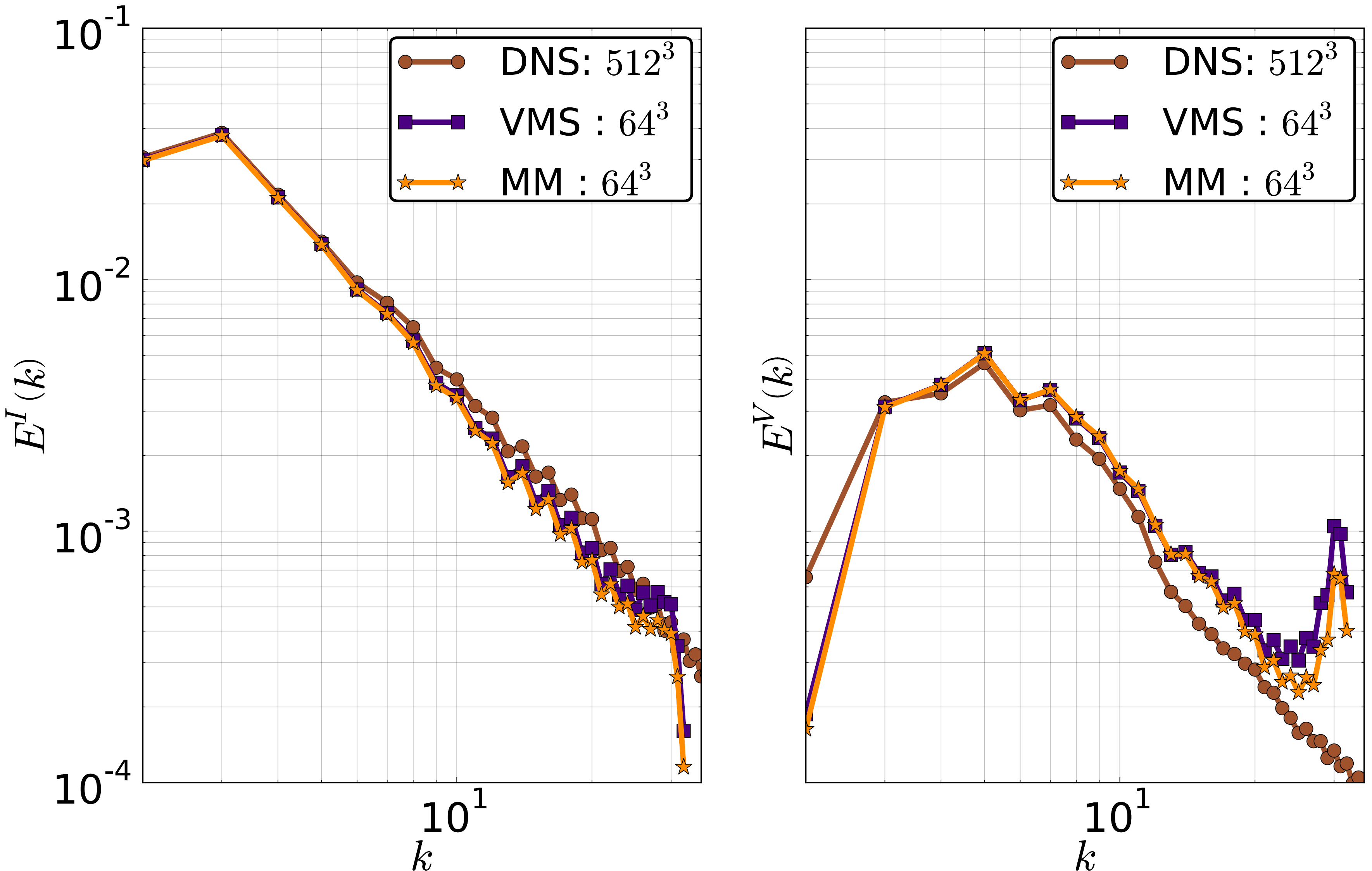}
      \caption{Magnetic and kinetic energy spectra at $t=4$.}
      \label{fig:EVEIofk}
\end{figure}

Next, we consider the total energy spectrum for relatively coarse FE discretizations.  Figure~\ref{fig:VMSvsMM} shows the total energy spectrum at $t=4$ for the VMS model and mixed model compared to the DNS results.  The LES results agree with the DNS results well for the large scales as expected.  Both models exhibit a slight pile-up of energy but this energy is dissipated in the last mode or two as it should be.  Further details of the solution behavior are uncovered in Figure~\ref{fig:EVEIofk} by considering the kinetic and magnetic energy spectra separately, also at $t=4$.  The DNS data for the magnetic energy spectrum is matched very well.  The models do not perform as well for coarse discretizations in replicating the kinetic energy.  Indeed, we see that the observed pile-up of total energy in the small scales stems directly from the kinetic energy.  This implies that the models do not perform as well for the momentum equation as for the induction equation.  This lack of performance is most likely due to highly active subgrid velocity scales.  The good agreement in the total energy spectrum is mainly due to the fact that this flow field is dominated by the magnetic field.  The energy in the velocity field is almost an order of magnitude less than that in the magnetic field.  An interesting point is that the performance discrepancy between the momentum and induction equation calls attention to the fact that we have selected our eddy viscosities such that the turbulent magnetic Prandtl number is unity.  It may be necessary to adjust the eddy viscosity portion of the model so that $\nu_{T}\neq\lambda_{T}$.

\begin{figure}[H]
  \centering
  \includegraphics[width=110mm,height=55mm]{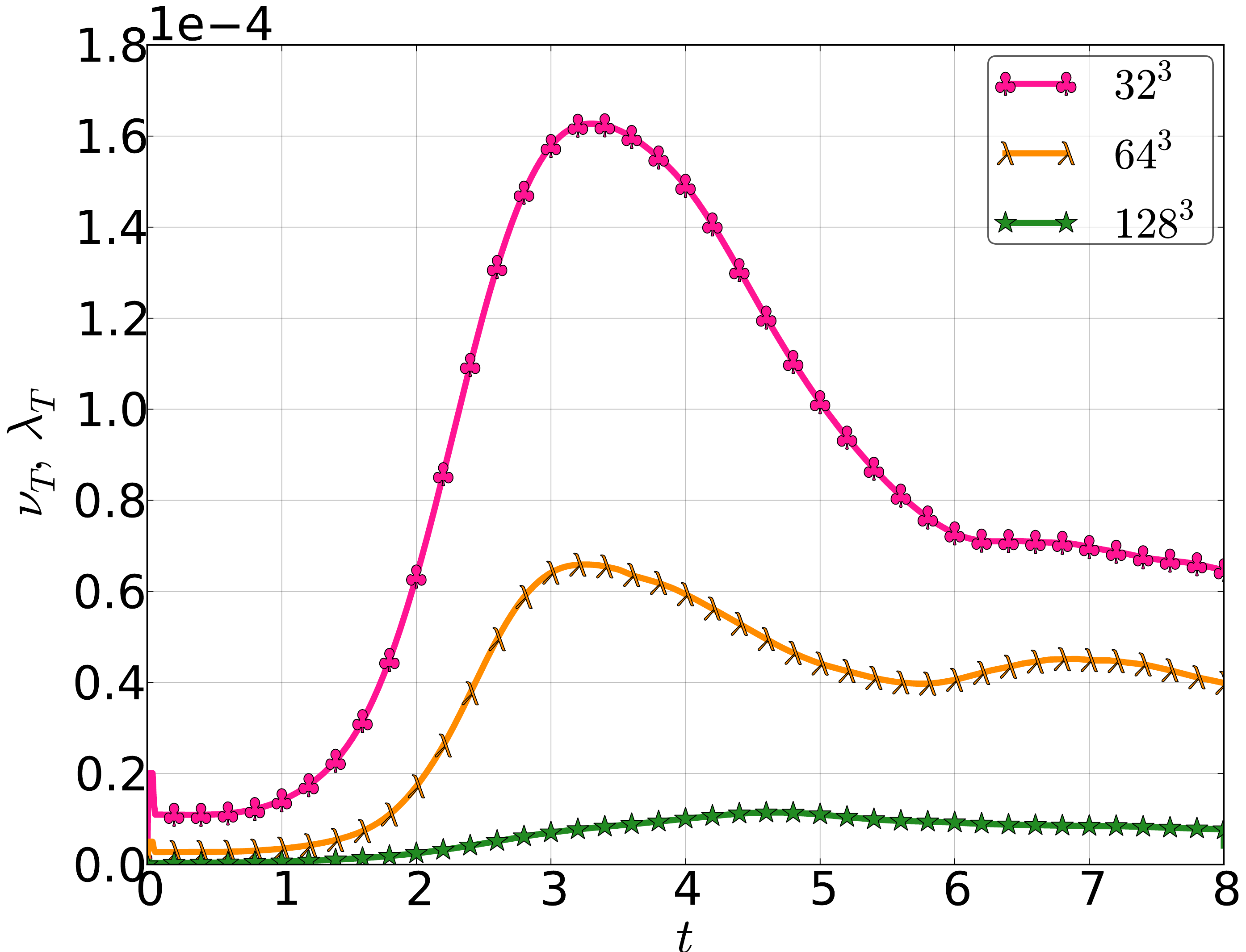}
  \caption{Time evolution of the MHD eddy viscosity at three different resolutions.  The inherent dynamic nature of the eddy viscosity is observed.}
  \label{fig:EV-evo}
\end{figure}

We can gain further insight into the nature and performance of the mixed model by inspecting the time evolution of the eddy viscosity.  In Figure~\ref{fig:EV-evo} we consider the time evolution of the cell-averaged eddy viscosity at three different spatial resolutions.  We observe that the eddy viscosity displays an automatic dynamic behavior.  Initially, the eddy viscosity is zero because there are no unresolved scales in the initial conditions and hence the residual is zero.  However, as soon as other scales become active the eddy viscosity ``turns on''.  It ramps up to a peak value at the peak of dissipation which is physically consistent; that is, the eddy viscosity should work hardest around the maximum of energy dissipation.  We also note that with spatial mesh-refinement the eddy viscosities become less active.  This is a direct result of the residual-dependence of the eddy viscosities; the numerical residual becomes smaller as the mesh is refined because the numerical solution is able to capture more scales of the flow field.  
Although encouraging, we also know from Figure~\ref{fig:VMSvsMM} that the eddy viscosity should be slightly larger in order to dissipate the energy from the small scales more efficiently.  Finally, we note that at the very beginning of the simulation the eddy viscosity exhibits a small overshoot.  This is the result of the finite element basis.  The initial condition is not contained in the finite element space.  Thus, the residual is immediately nonzero.  In a Fourier spectral method of the Taylor-Green vortex in hydrodynamics this overshoot is not present~\cite{oberai2014residual}.

\begin{figure}[H]
  \centering
  \includegraphics[width=110mm,height=65mm]{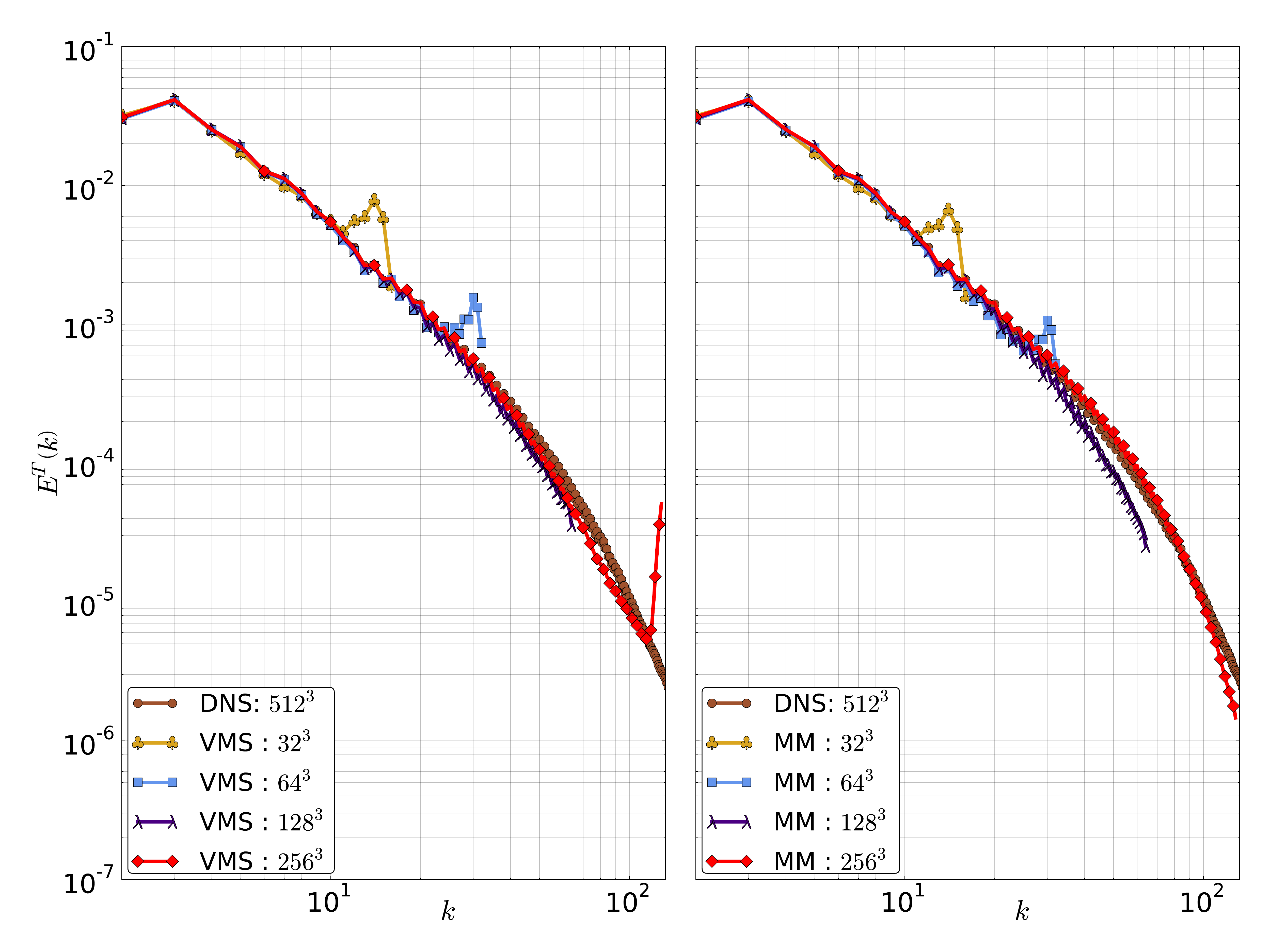}
  \caption{Convergence of total MHD energy spectrum with mesh refinement at $t=4$.}
  \label{fig:ET4s_mc}
\end{figure}

\begin{figure}[H]
  \centering
  \includegraphics[width=110mm,height=65mm]{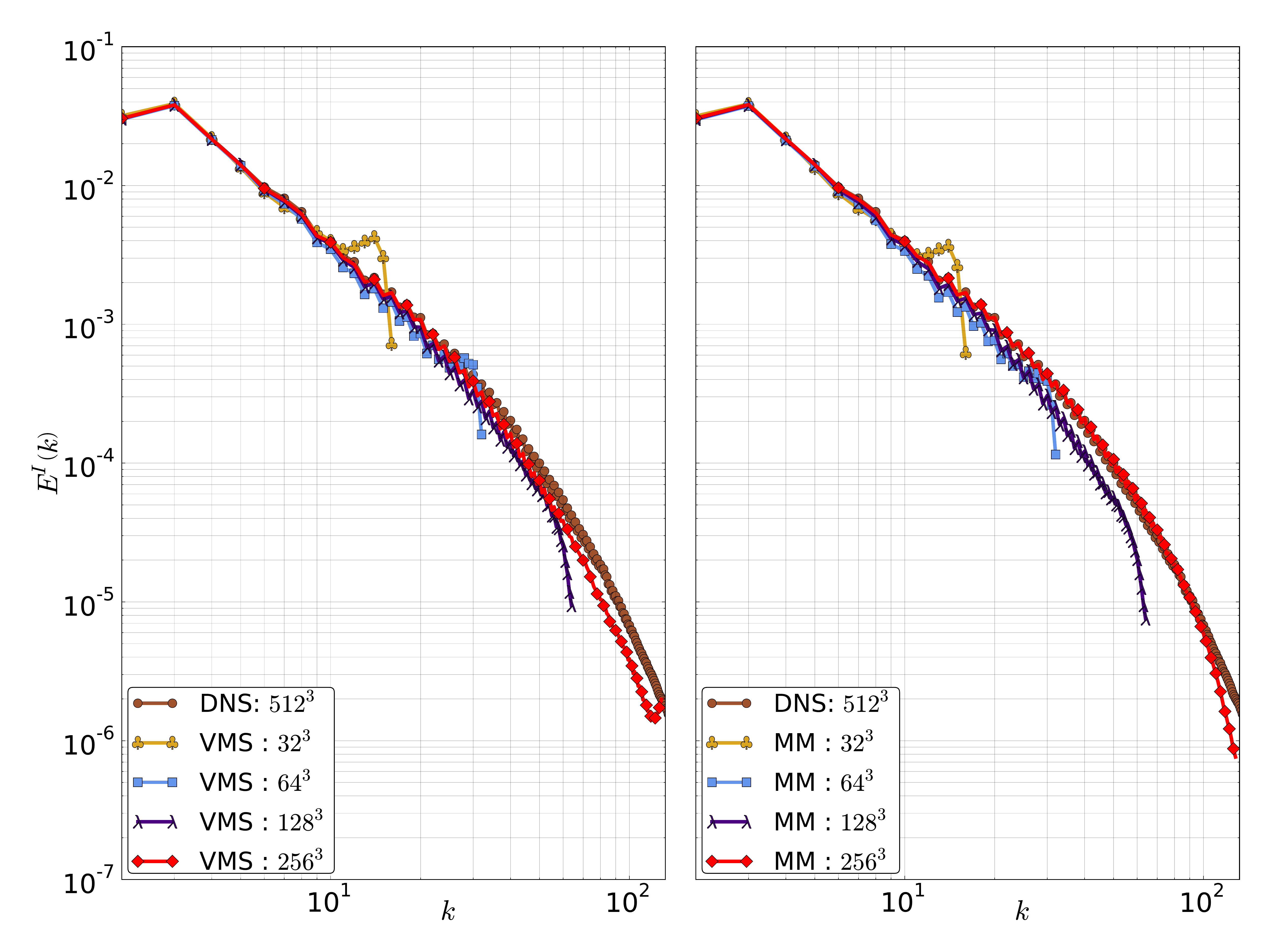}
  \caption{Convergence of magnetic energy spectrum with mesh refinement at $t=4$.}
  \label{fig:EI4s_mc}
\end{figure}

\begin{figure}[H]
  \centering  
  \includegraphics[width=110mm,height=65mm]{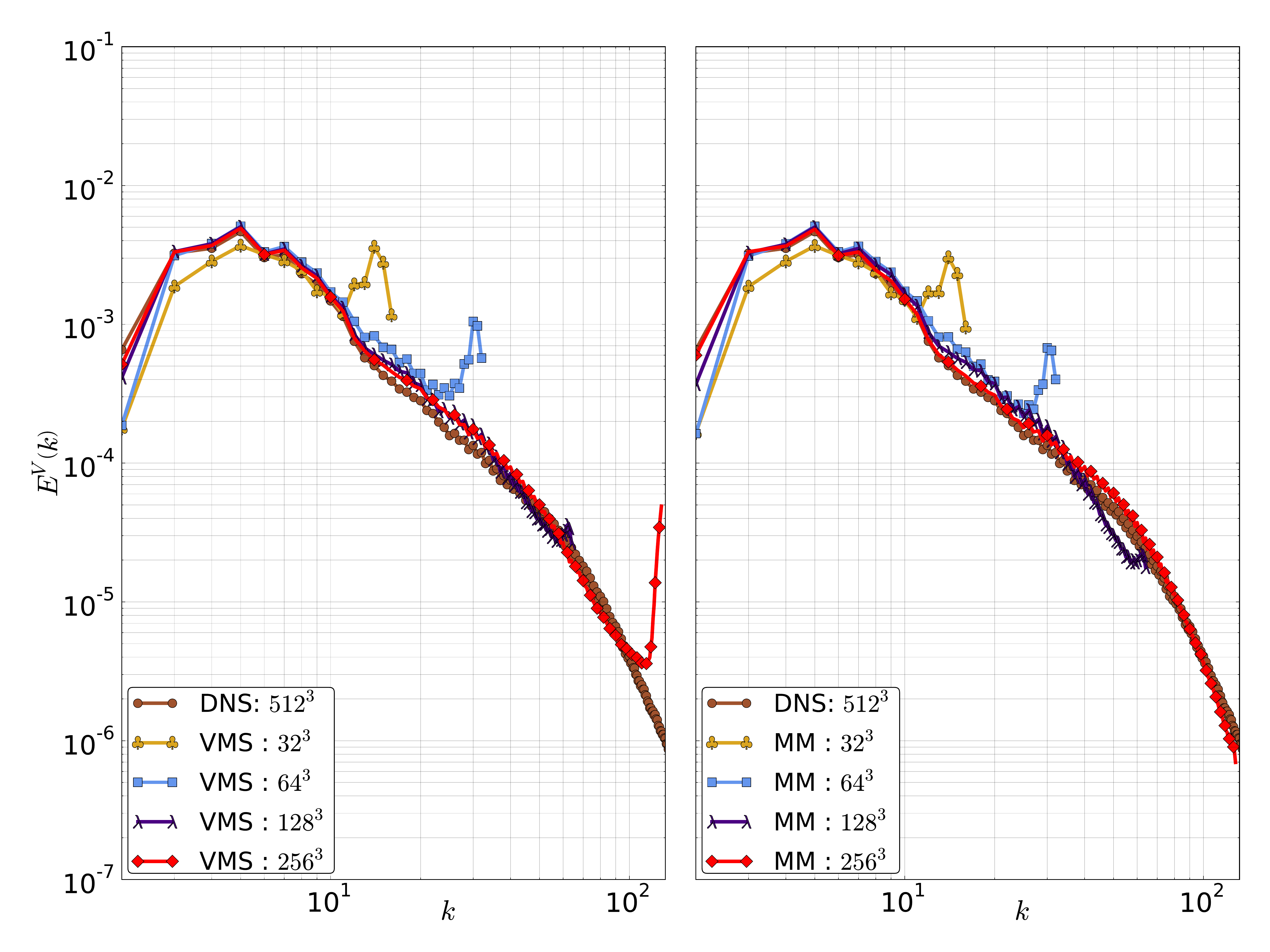}
  \caption{Convergence of kinetic energy spectrum with mesh refinement at $t=4$.}
  \label{fig:EV4s_mc}
\end{figure}

Figures~\ref{fig:ET4s_mc}-\ref{fig:EV4s_mc} show energy spectra at various resolutions at $t=4$.  
In all cases, the LES results converge to the DNS results.  At $N=128$ and $N=256$ the LES results are quite good.  We note that for the VMS model there is a pile-up of energy in the last few modes of the kinetic energy spectrum at $N=256$.  However, this pile up is $\mathcal{O}\lr{10^{-5}}$ and is quite small.  It is most likely due to the fact that the subgrid scales at such high wavenumbers are very small and the models have some difficulty accurately representing such small scales.  The convergence in the case of the mixed model is seen to be much better with the high wavenumber modes being controlled much more effectively.  The correspondence of the $N=256$ MM discretization with the 512 mode spectral DNS solution shows the remarkable ability of this model to reproduce the spectrum accurately. 

%% file: HighReResults.tex
\subsection{High Reynolds Number}\label{sec:highReresults}
\begin{figure}[H]
  \centering
  \includegraphics[width=110mm,height=70mm]{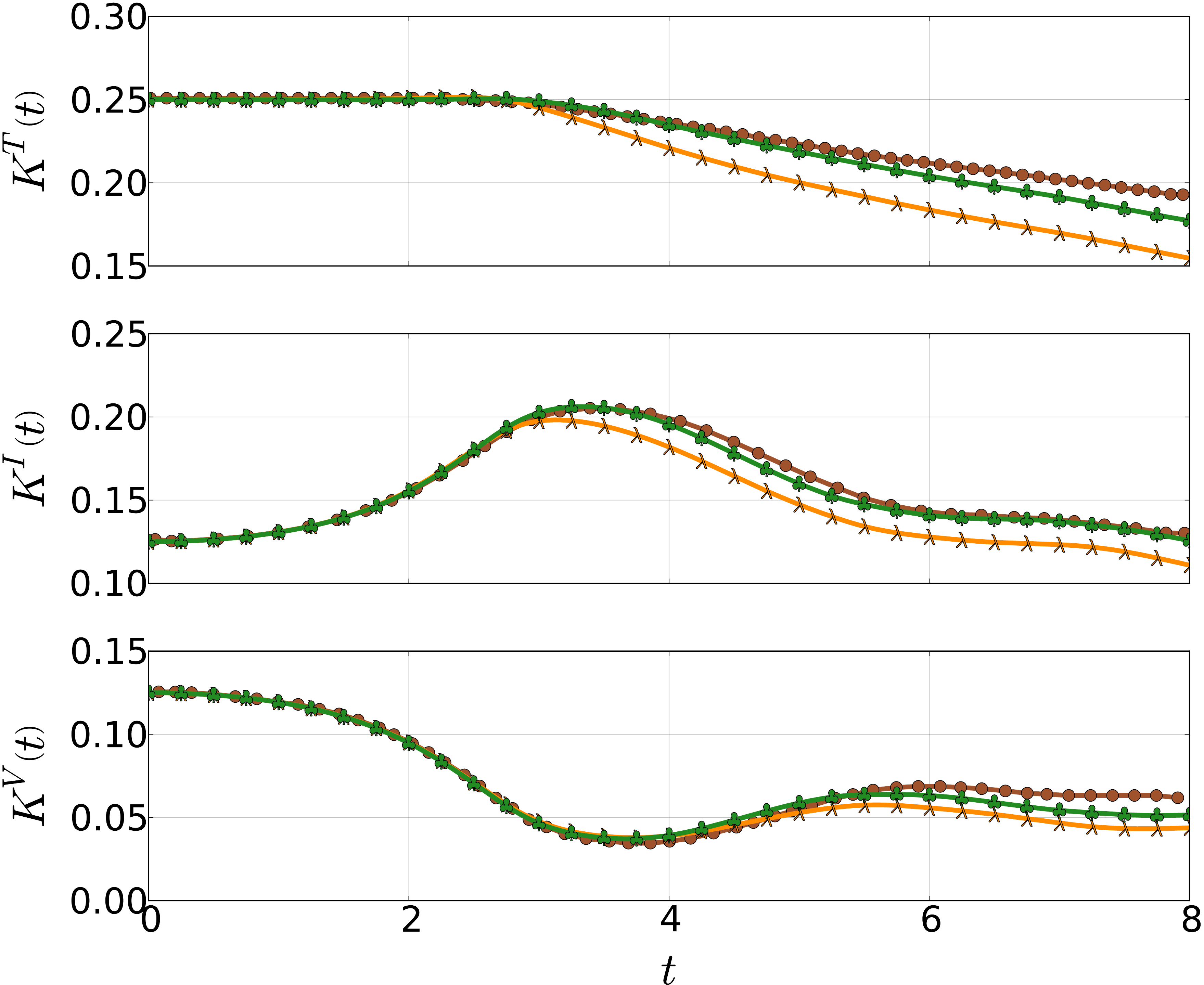}
  \caption{Comparison of time evolution of total, magnetic, and kinetic energies between DNS results (brown circles) and mixed model results at resolutions of $128^3$ (orange lambdas) and $256^3$ (green clubs).}
  \label{fig:HRe_Koft}
\end{figure}

\begin{figure}[H]
     \centering
     \includegraphics[width=110mm,height=65mm]{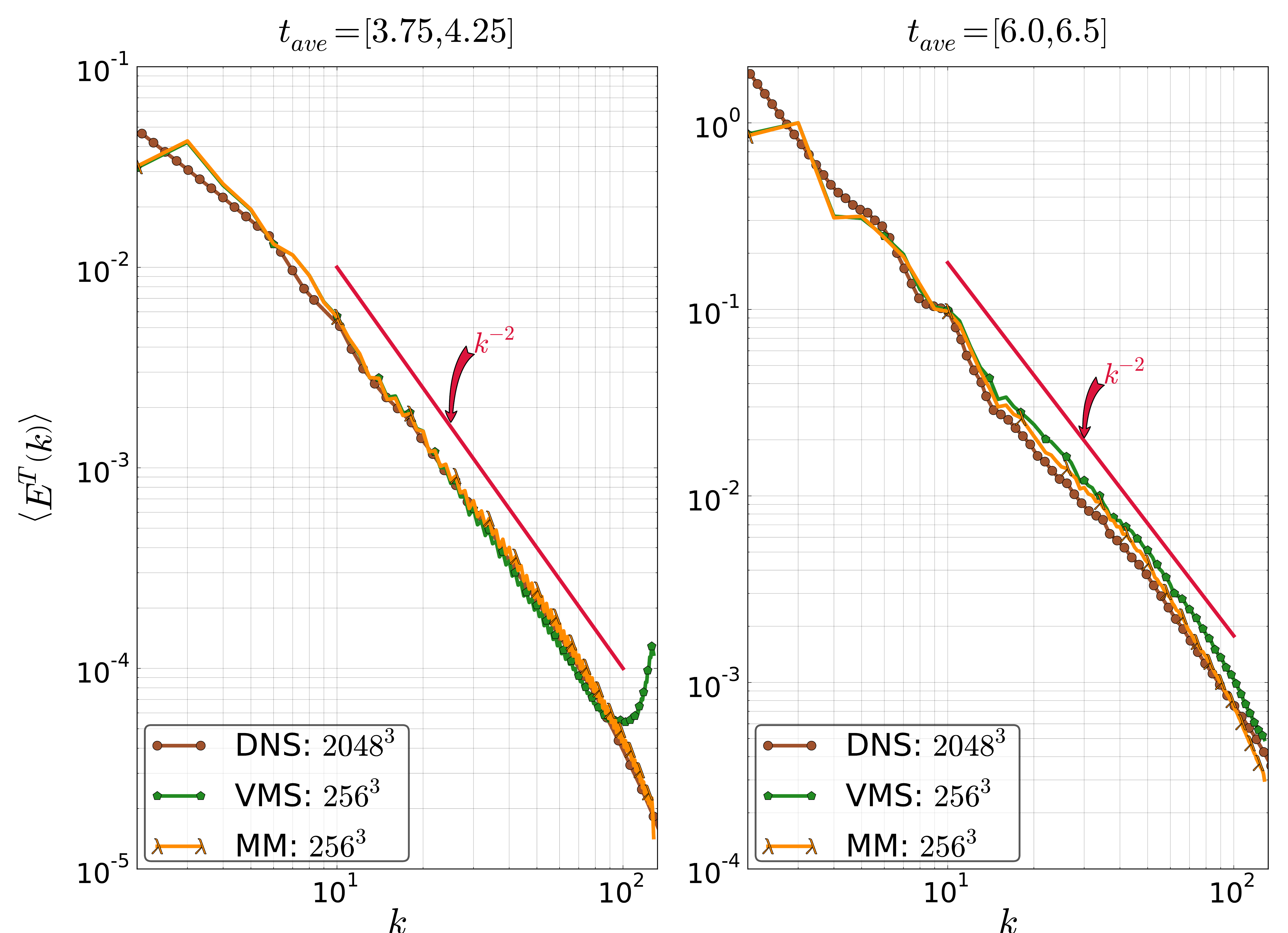}
     \caption{Performance of VMS and mixed models for high Reynolds numbers at two different times.  The total energy spectra have been averaged around the times indicated in the figure.}
     \label{fig:high_Re_VMS_MM}
\end{figure}

\begin{figure}[H]
     \centering
     \includegraphics[width=110mm,height=55mm]{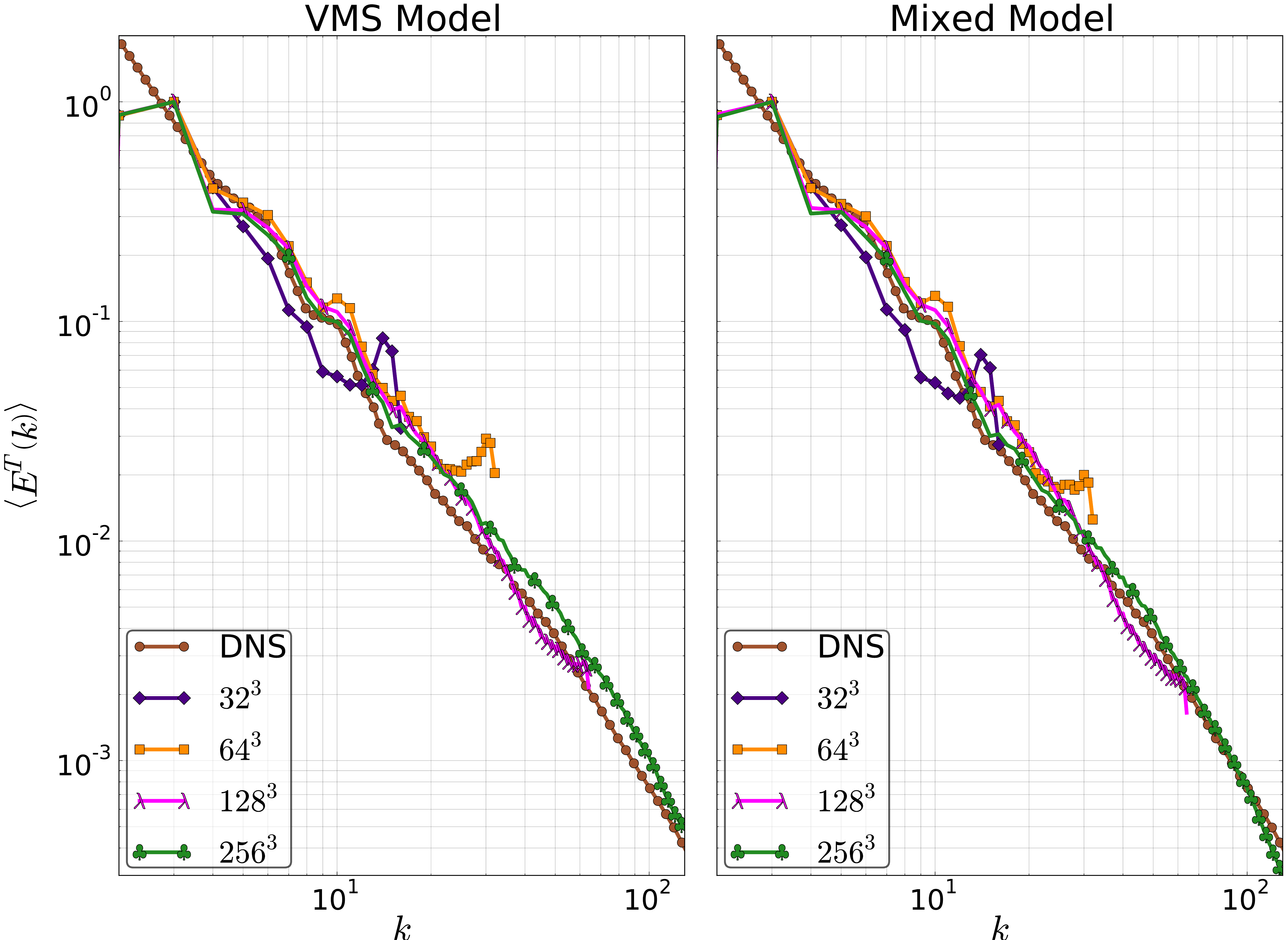}
     \caption{Convergence of FEM VMS and mixed models demonstrated using $32^3$ - $256^3$ modes.  The spectra were averaged around $t\in\left[6, \ 6.5 \right]$}
     \label{fig:high_Re_convergence}
\end{figure}

Next we turn our attention to the same flow field at higher Reynolds numbers.  The Reynolds numbers near the peak of dissipation are $Re=Rm=9000$.  Figure~\ref{fig:HRe_Koft} presents the time evolution of energies obtained from DNS results and the mixed model at two different resolutions.  The results are quite good especially for $N=256$ for which the LES solutions are only slightly overly dissipative for later times.
We also consider, once again, plots of the total energy spectrum and compare the spectra from the models to DNS spectra from~\cite{lee2010lack} in Figures~\ref{fig:high_Re_VMS_MM} and~\ref{fig:high_Re_convergence}.  This time, however, the spectra have been averaged over certain time intervals which are indicated in the respective figures.  We have performed LES runs at various spatial resolutions at $N=32, \ 64, \ 128, \ 256$.  Note that even at the highest LES resolution the DNS result contains $8^3$ times as many modes in the expansion compared to the number of elements in the LES results. Given this difference in resolution, and the exponential convergence of the spectral method the correspondence is remarkable. Further, the ability of the reasonably coarse LES type models to approximate the spectrum is encouraging in the context of controlling the computational costs of such simulations. Figure~\ref{fig:high_Re_VMS_MM} compares the performance of the LES models at $N=256$ to the DNS results.  We observe that both the VMS and mixed models perform quite well in the inertial range at both times.  However, when the spectrum is averaged around the peak of dissipation (leftmost plot in Figure~\ref{fig:high_Re_VMS_MM}) we notice that the VMS model results in some energy pile-up in the smallest scales.  We note that the energy pile-up is on the order of $10^{-5}$ and is therefore rather small.  Morever, in the last one or two modes, the VMS model begins to disspate energy, albeit inadequately.  On the other hand, the mixed model performs admirably and matches the DNS spectrum nearly identically.  The only difference between the mixed model and VMS model is that the mixed model attempts to account for the second order correlations that the VMS model neglected.  For high Reynolds numbers these second order correlations are expected to be significant.  One would expect that near the peak of dissipation these correlations are particularly active.  This explains why the spectrum from the VMS model exhibits some energy pile-up while the mixed model dissipates energy at precisely the correct rate.

The rightmost plot in Figure~\ref{fig:high_Re_VMS_MM} shows the spectra averaged around a later time that is not near the peak of dissipation.  We observe that the VMS model and mixed model both perform well but that the mixed model performs slightly better.  The biggest difference between this plot and the similar plot near the peak of dissipation is that the VMS model does not result in a pile-up of energy in the smallest scales.  This makes sense in light of the fact that the dissipation rate is not as intense and that the higher order correlations are therefore less active.  In addition, we note that the LES results have the correct inertial range scaling of $k^{-2}$.  This is particularly interesting because in the derivation of the mixed model constant we \textit{assumed} that the energy spectrum would follow a Kolmogorov scaling law.  The good results are a further indication of the dynamic nature of the eddy viscosity portion of this model.  That is, the model adapts as necessary to the dynamics of the flow field.  The exact value of the mixed model constant does not appear to be crucial.

Finally, Figure~\ref{fig:high_Re_convergence} shows plots of the VMS and mixed models at various spatial resolutions in order to demonstrate qualitative convergence with mesh refinement.  We observe that the $32^3$ simulations are excessively dissipative as expected from linear finite elements.  However, with subsequent refinements the spectrum matches the DNS better, especially in the inertial range.  Each refinement shows improvement in agreement between the LES models and the DNS result. Clearly the correspondence of the spectrum for the mixed model is very impressive and indicates that this MHD LES model is performing very well.

%% file: Conclusions.tex
\section{Conclusions}\label{sec:conclusions}

In this work, we have introduced a class of finite element turbulence models for incompressible magnetohydrodynamics.  These models are derived from the variational multiscale formulation wherein the fine scales are assumed to be directly proportional to the residual of the coarse scales.  The general expression for the models is provided in~\eqref{eq:all_models} and specific model choices are summarized in Table~\ref{tab:b_parameters}.  We briefly summarize the new models:
\begin{enumerate}
  \item The variational multiscale formulation is given by~\eqref{eq:vms_implemented} where the fine scales are determined from~\eqref{eq:uprime}.
  \item Eddy viscosity models are given by~\eqref{eq:var_model} and~\eqref{eq:mhd_evm}.  Different choices of the eddy diffusivites $\nu_{T}$ and $\lambda_{T}$ result in different models.  Dynamic Smagorinsky models would result from selecting either~\eqref{eq:nut_dsev} and~\eqref{eq:lambdat_dsev} for $\nu_{T}$ and $\lambda_{T}$ respectively while the alignment-based dynamic Smagorinsky model would result from~\eqref{eq:nut_aligned} and~\eqref{eq:lambdat_aligned}.  We have introduced an alternative eddy viscosity model which we refer to as the residual-based eddy viscosity model.  In the RBEV model, the eddy diffusivities are proprotional to the unresolved velocity and magnetic field as determined via the VMS method.  In this way, the eddy viscosity model is fully residual based and therefore results in a consistent numerical method.  Moreover, we have assumed that the constant appearing in this model does not need to be determined dynamically.  The RBEV model uses~\eqref{eq:rbev_diff} for the eddy viscosities.
  \item The mixed model is given by~\eqref{eq:mixed_model}.  In this work we have used the VMS models along with the RBEV model.
\end{enumerate}

The new models were tested on the Taylor-Green vortex generalized to MHD at two different Reynolds numbers.  Comparisons were made to existing DNS data in terms of the evolution of energies as well as energy spectra at various times.  The mixed model slightly outperformed the pure VMS model as expected.  All models struggled slightly in reproducing the DNS energy spectra at the peak of dissipation except for the $N=256$, high Reynolds number case in which the mixed model dissipated energy at the correct rate and matched the DNS data very well.  The models were able to eventually dissipate the energy even in cases where slight energy pile-up was observed.  It is interesting to note that the models were able to capture the correct inertial range behavior for this flow field even though it does not display a traditional Kolmogorov energy spectrum.  This is in spite of the fact that the RBEV model consant was derived assuming a Kolmogorov spectrum.  This behavior is not entirely surprising, however, considering that the eddy viscosity models are inherently dynamic and adjust automatically to the flow field.

Certain numerical properties of the new models were explored in addition to physical considerations.  In particular, the nature of the eddy viscosities was considered as well as convergence properties of the method.  The eddy viscosities displayed their expected dynamic behavior, peaking at the maximum of physical dissipation and decreasing thereafter.  With mesh refinement, the eddy viscosities decreased in strength because the role of the subgrid scales was diminished.  The numerical solution also improved with mesh refinement.

Considering the encouraging performance of the new proposed models, nevertheless, there are still several potential areas of improvement.  The RBEV constant could be improved although it does not seem likely that a true universal constant for MHD can be found due to the nonuniversality of decaying MHD turbulence.  This lack of universality may not be a significant hindrance to these models however due to their dynamic nature.  Already, as shown in this work, the models are able to capture the correct inertial range behavior.  The RBEV constant was derived in a spectral context and improvements to this constant may involve considering alternative derivations in terms of finite elements.  In general, the models appear to be quite sensitive to their constant parameters.  This is especially true in the definition of the stabilization matrix found in~\eqref{eq:tau_mom} -~\eqref{eq:tau_r}.  A parameter sensitivity analysis may shed some light on the role that these constants play in affecting the solutions to different problems.

The VMS method is not guaranteed to be globally dissipative; this feature depends on the design of the finite element function spaces.  The models developed in the present work do not guarantee this global dissipativity and this fact is seen in the coarse discretization plots of the time evolution of total MHD energy in Figure~\ref{fig:KeT}.  Namely, just prior to the peak of dissipation, the total MHD energy slightly exceeds the initial energy.  A careful consideration of the finite element spaces as is done in~\cite{codina2002stabilized} may help to overcome this issue.

There are several avenues left to explore in the future related to physical and numerical considerations.  The optimal blending of parameters in the general mixed model presented in~\eqref{eq:all_models} is still undetermined.  For now, all of the models appearing in~\eqref{eq:all_models} are weighted equally.  Some evidence exists that the eddy viscosity portion of the complete model should be weighted less than the VMS portion.  This is based upon arguments concerning the contribution of cross stresses and Reynolds stresses in hydrodynamic turbulence~\cite{kraichnan1976eddy}.

One nice feature of the VMS-based models is the fact that they permit local backscatter.  In MHD turbulence this feature is quite important as it permits the transfer of energy from the turbulent velocity fluctuations into the large scale magnetic field.  This phenomenon has implications in models of the geodynamo.  We have begun studies to quantify this phenomenon in relation to the turbulence models developed in this work both in a spectral and finite element setting.  Future work will also include the application of these models to complex geometries that include modeling liquid metals flows for fusion applications~\cite{smolentsev2010mhd} and geodynamo simulations~\cite{kono2002recent}.

%% file: Acknowledgements.tex
\section*{Acknowledgements}\label{sec:ack}
This work was initiated with support from the Department of Energy (DOE) Office of Science Graduate Fellowship (SCGF) under Contract No. DE-AC05-06OR23100.  Support from NSF-DMS grant 1147523 is gratefully acknoweledged during preparation of this paper. Additionally the work of Shadid, Pawlowski, Cyr and Smith was partially supported by
the DOE Office of Science Applied Mathematics Program at Sandia National Laboratories  under contract DE-AC04-94AL85000.